\pdfoutput=1 
\documentclass[prl,groupedaddress]{revtex4}
\newcommand{\bra}[1]{\langle #1 |  }
\newcommand{\ket}[1]{|  #1 \rangle }
\newcommand{\Tr}[1]{  {\rm Tr} \left\{ #1 \right\} }

\usepackage[pdftex]{graphicx}

\title{} 
\date{}

\begin{document}


\begin{center}
{\huge{Energy conversion in Purple Bacteria Photosynthesis}}
\end{center}
\begin{flushright}

{\small{Chapter to appear in {\it Photosynthesis}, INTECH, 2011}}\\
 \vspace{1 cm}
 
Felipe Caycedo-Soler$^{1,2}$, Ferney J. Rodr\'iguez$^2$, Luis Quiroga$^2$,\\ Guannan Zhao$^3$ and Neil F. Johnson$^3$\\

{\it $^1$Ulm University, Institute of Theoretical Physics, Ulm,Germany\\
$^{2}$Departamento de F\'{i}sica, Universidad de los Andes,  Bogot\'a, Colombia\\
$^{3}$Department of Physics, University of Miami, Coral Gables, Miami, Florida, USA}
\end{flushright}

\section*{Introduction }
The study of how photosynthetic organisms convert light offers insight not only into nature's evolutionary process, but may also give clues as to how best to design and manipulate artificial photosynthetic systems  -- and also how far we can drive natural photosynthetic systems beyond normal operating conditions, so that they can harvest energy for us under otherwise extreme conditions. In addition to its interest from a basic scientific perspective, therefore, the goal to develop a deep quantitative understanding of photosynthesis offers the potential payoff of enhancing our current arsenal of alternative energy sources for the future.

In the following Chapter, we consider the excitation dynamics of photosynthetic membranes in {\it Rps. Photometricum} purple bacteria.
The first studies on purple bacteria photosynthetic membranes were concerned with the complex underlying  detailed structure \citep{mcdermott95,roszack2003,waltz1998,jamieson2002}. The interested reader might find helpful the first section of this chapter where we present a summary of structures and  processes that allow photosynthesis in  purple bacteria. As improved resolution became available for light-harvesting structures, so too the community's interest increased in  understanding the details of the rapid femto- to picosecond timescales for the excitation transfer process within a given harvesting complex -- even to the regime where quantum effects are expected. Indeed such quantum effects have recently been confirmed in, for instance, the Fenna-Matthews-Olson complex of green sulfur bacteria \citep{FMOnature}.  However, the processes occurring at this level of detail in terms of both structure and properties of the excitation harvesting, have not yet been shown as being crucial to the performance of the full harvesting membrane, nor with the primary goal of any given photosynthetic organism: to fuel its metabolism.
We focus on the transfer among different complexes, in particular the inter-complex excitation transfer. We consider first a system of few complexes in order to understand the consequences on the dynamics of the complexes' connectedness. We also consider the relative amounts of harvesting complexes (i.e. stoichiometry) on small sized networks in order to help establish our understanding of the behavior of complete chromatophore vesicles.

As a whole, chromatophore vesicles comprise sections of the purple bacteria cytoplasmic membrane where a large number of harvesting complexes accommodate. The current capabilities to dissect these vesicles on the nanoscale through Atomic Force Microscopy (AFM) has provided evidence of changes in the conformation of chromatophores as a result of different environmental conditions. According to Ref.\citep{scheuring2005} , membranes grown under Low Light Intensity (LLI,  10 Watt/m$^2$) present a relative amount of harvesting complexes which is different to the stoichiometry observed for bacteria  grown under High Light Intensity (HLI, 100 Watt/m$^2$), thereby pinpointing the importance of global changes in the complete vesicles as an important means of fulfilling the bacteria's metabolic requirements.  In order to study these global conformational changes -- given the fact that the inter-complex transfer time-scale involves several picoseconds and the excitation delocalization length is not expected to be beyond a single harvesting complex -- we choose a model of excitation dynamics which is based on a classical random walk. This random walk is coupled to the main processes leading to electron/chemical energy transformation, and hence the bacteria's metabolic demands.

This Chapter is organized as follows. Section 1 provides a review of the basic structures involved in excitation transfer, along with a summary of the processes required for electron/chemical energy conversion. A discussion of the dynamics of excitations in a few model architectures is explored in section 2,  in order to understand
the results from complete LLI  and HLI adapted chromatophores in section 3 and to guide our development of an analytical model in section 4 for determining both the efficiency and power output of any given chromatophore vesicle under arbitrary light intensity regimes. Lastly, in section 5 we explore the effect of incident light with extreme photon arrival statistics, on the resulting electronic/chemical energy conversion, in order to heuristically provide a survivability margin beyond which terrestrial bacteria could not survive. This is motivated by the fact that one day, it may be necessary to send simple bacteria into deep space and/or cope with extreme photon conditions here on Earth as a result of a catastrophic solar change.


\section*{1. Important processes for solar/chemical energy conversion}\label{structure}

\begin{figure}
\centering
\includegraphics[width=12 cm]{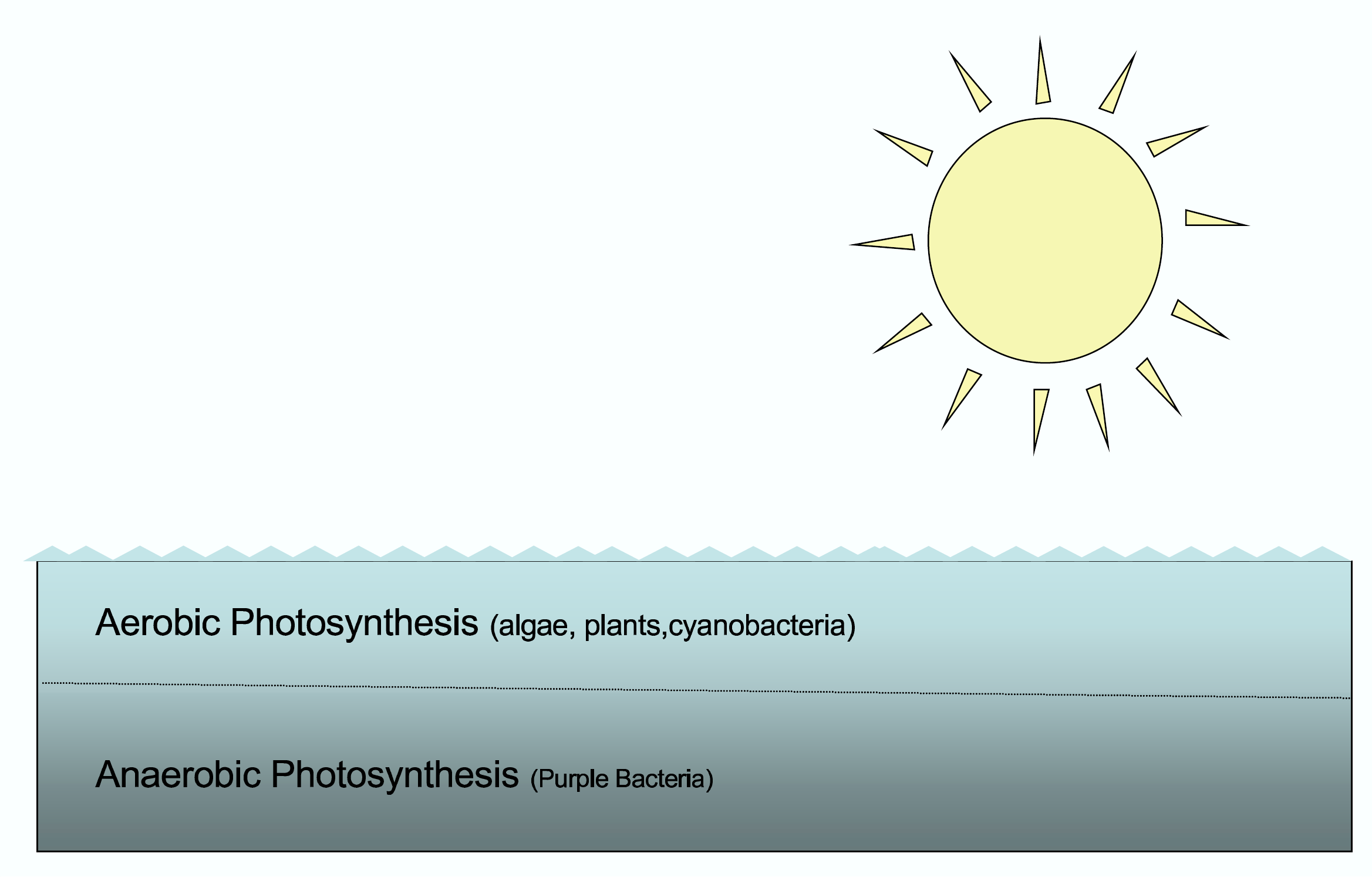}
\caption{Representation of a lake containing both aerobic and anaerobic phototrophic organisms. Note that purple bacterial photosynthesis is restricted to the lower anaerobic layer and so they only receive solar energy that has been filtered, mainly by chlorophylls belonging to algae, cyanobacteria and plants. }\label{fig1}
\end{figure}

Purple bacteria sustain their metabolism using photosynthesis in anaerobic conditions and under the dim light excitation proper of several meters deeps at ponds, lagoons and streams  \citep{Pfenning}. As depicted in Fig.\ref{fig1}, aerobic organisms are present near the surface of water reservoirs, collect blue and red  spectral components of sun's light, leaving only the green and far red ($>750 $nm) components, from where purple bacteria must fulfill their energy requirements.

\begin{figure}
\hspace{-1.5 cm}
\vspace{-0 cm}
\includegraphics[width=16 cm]{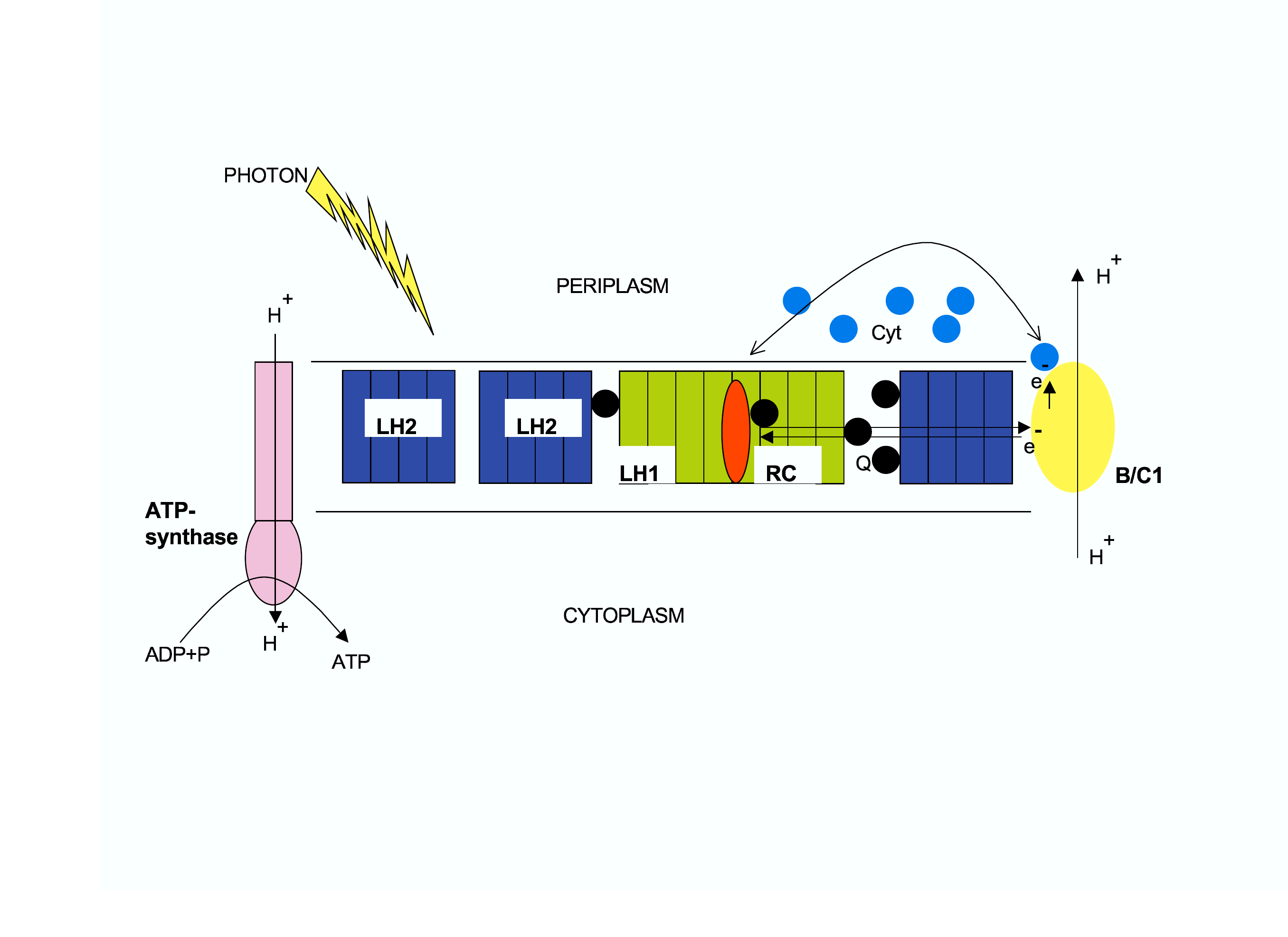}
\vspace{-3 cm}
\caption{Schematic representation of the photosynthetic apparatus in the intracytoplasmic membrane of purple bacteria. The reaction center (RC, red) is surrounded by the light-harvesting complex 1 (LH1, green) to form the LH1+RC  complex, which is surrounded by multiple light-harvesting complexes (LH2s) (blue), forming altogether the photosynthetic unit (PSU). Photons are absorbed by LHs and excitation is transferred to the RC initiating a charge (electron-hole) separation. Electrons are shuttled back by the cytochrome $c_2$ charge carrier (blue) from the ubiquinone-cytochrome $bc_1$ complex (yellow) to the RC. The electron transfer across the membrane produces a  proton gradient that drives the synthesis of ATP from ADP by the ATPase (orange). Electron $e^-$ is represented in blue, and quinones $Q_B$, likely confined to the intramembrane space, in black.}\label{fig2}
\end{figure}

The light energy absorption is accomplished through intracytoplasmic  membranes where different pigment-protein complexes accommodate. Light Harvesting complexes (LHs) have the function of absorbing light and transfer it to Reaction Centers (RCs), where a charge separation process is initiated \citep{codgell}. The unpaired charge reduces a quinone, which using a periplasmic hydrogen, converts to quinol (Q$_B$H$_2$). RCs neutrality is restablished thanks to cytochrome $cyt$ charge carrier, which after undocking from the RC, must find the $bc_1$ complex to receive an electron and start its cycle all over again. The electron in $bc_1$ is given due to cytoplasmic Quinol delivery in $bc_1$. The proton gradient induced by the charge carriers cycling becomes the precursor of adenosyn triphosphate (ATP) synthesis: ADP+P$\rightarrow$ ATP+Energy, where ADP and P refer respectively, to adenosine diphosphate and phosphorous. The cycle is depicted in Fig.\ref{fig2}.

\subsection*{1.1 Harvesting function}
Light absorption occurs through organic molecules, known as chromophores, inserted in protein complexes. Bacteriochlorophyls (BChl) and carotenoids (Car) chromophores  are the main absorbers in purple bacteria photosynthesis,  principally in the far red and green, respectively. The light absorption process occurs through chromophore's $Qy$ electronic transition excitation. Several chromophores are embedded in  protein helices, named $\alpha$ and $\beta$ apoproteins, inside complexes, classified by their absorption spectral maximum. Light Harvesting complex 2 (LH2) reveals two concentric subunits that according to their absorption maxima are called B800 or B850, composed of nine pairs of apoproteins \citep{mcdermott95}  comprising an inner $\alpha$-helix and an outer $\beta$-helix both crossing the harvesting membrane from periplasm to cytoplasm,  in an $\alpha\beta$ unit that serves to anchor a highly interacting B850 dimer and one B800 chromophore. Hence, the B800 is composed of nine chromophores, while B850 include eighteen BChl{\it a} chromophores, having dipole moments parallel and  nearly perpendicular to the membrane plane, respectively. Raman spectra using different excitation wavelength \citep{gall2006} and stoichiometry analysis \citep{arellano98}, indicated that one carotenoid (Car) is present per each $\alpha\beta$ unit.  Light Harvesting complex 1 (LH1) absorbs maximally at 883 nm, and contain 32 BChls, arranged in 16 bi-chromophore $\alpha\beta$  units, surrounding an RC \citep{karrasch95} in the same geometrical arrangement as B850 chromophores. An RC presents a highly interacting dimer, the special pair (P), that is ionized due to the electronic excitation transferred from the surrounding LH1.

The complexes' photon absorption cross section has been calculated for LH1 and LH2 complexes, where all absorbing molecules and extinction coefficients \citep{francke95} have been taken into account.  A photon of wavelentht $\lambda$, is part of the power spectrum of a source  with occupation numbers $n(\lambda)$.  Normalized to 18 W/m$^2$ intensity, the rate of photon absorption  for circular LH1 complexes in {\it Rb. sphaeroides} \citep{geyer2006}:
\begin{equation}
\gamma^A_{1}=\int n(\lambda) \sigma_{\mbox{LH1}}(\lambda)d\lambda=18 \mbox{s}^{-1}.
\end{equation}
The same procedure applied to LH2 complexes, yields a photon capture rate of $\gamma^A_2=10 $s$^{-1}$. Since these rates are normalized to 18 W/m$^2$, the extension to arbitrary light intensity $I$ is straightforward. The rate of photon absorption normalized to 1 W/m$^2$ intensity, will be $\gamma_{1(2)}=1 (0.55) $s$^{-1}$. From now on, subindexes 1 and 2 relate to quantities of LH1 and LH2 complexes, respectively.  Vesicles containing several hundreds of complexes will have an absorption rate:
\begin{equation}\label{gammaa}
\gamma_A=I(\gamma_1 N_1+\gamma_2N_2)
\end{equation}
where $ N_{1(2)}$ is the number of LH1 (LH2) complexes in the vesicle.

\subsection*{1.2 Excitation transfer}\label{exctans}
Excitation transfer happens through Coulomb interaction of electrons, excited to the  Q$_y$ electronic transition in chromophores. The interaction energy can be formally written \citep{fcondon}:
\begin{equation}\label{vij}
V_{ij}=\frac{1}{2}\sum_{m,n,p,q}\sum_{\sigma,\sigma'}\bra{\phi_m\phi_n}V\ket{\phi_p\phi_q} c^+_{m\sigma}c^+_{p\sigma'}c_{q\sigma}c_{n\sigma},
\end{equation}
where $c^+_{m\sigma}$, $c_{n\sigma'}$ are fermion creation and annihilation operators of electrons with spin $\sigma$ and $\sigma'$, in the mutually orthogonal atomic orbitals $\phi_m$ and $\phi_n$.
 The overlap $ \bra{\phi_m\phi_n}V\ket{\phi_p\phi_q}$ is the Coulomb integral:
\begin{equation}\label{vda}
\bra{\phi_m\phi_p}V\ket{\phi_n\phi_q}=\int\int d\vec{r}_1\,d\vec{r_2} \, \phi_m^*(\vec{r_1}-\vec{r_i})\phi_p(\vec{r_1}-\vec{r_i})\frac{e^2}{\vert \vec{r_1}-\vec{r_2}\vert}\phi_n^*(\vec{r_2}-\vec{r_j})\phi_q(\vec{r_2}-\vec{r_j})
\end{equation}
that accounts on inter-molecular exchange contribution when donor and acceptor are at a distance comparable to the extent of the molecules, and  the direct Coulomb  contribution for an electron that makes a transition between $\phi_m$ and $\phi_p$, having both a finite value near the position of the donnor $\vec{r_1}\approx \vec{r}_D$, while another electron is excited between $\phi_n$ and $\phi_q$ at the acceptor coordinate $\vec{r_2}\approx \vec{r}_A$. In this latter situation, a commonly used  framework concerns a tight-binding Hamiltonian, where details in the specific molecular orbitals $\ket{\phi_n}$ involving mainly the $Q_y$ orbitals is left aside, and emphasis relies on occupation with a single index labeling an electronic state $\ket{i}$ concerning occupation in a given chromophore. The Hamiltonian $H$ in the chromophore site basis $\vert i\rangle$,
\begin{equation}
H=\epsilon\sum_i\vert i\rangle\langle i\vert +\sum_{i,j}V_{ij}\vert i\rangle \langle j\vert
\end{equation}
has diagonal elements $\langle i\vert H \vert i\rangle=\epsilon$, concerning the energy of the excitation, usually measured  from the ground electronic state.  Neighboring chromophores  are too close to neglect their charge distribution and its interaction is determined such that the effective Hamiltonian spectrum matches the spectrum of an extensive quantum chemistry calculation \citep{hu1997}. In the LH2 complex, the B850 ring, with nearest neighbors coupling  $\langle i\vert H\vert i+1\rangle= 806$ cm$^{-1}$  or 377 cm $^{-1}$ for chromophores  respectively, within or in different  neighboring $\alpha\beta$ units.  For next-to-neighboring chromophores the dipole-dipole approximation is usually used,
\begin{eqnarray}
\langle i \vert H\vert j\rangle &=&  V_{ij}=C\left(\frac{\vec{\mu}_i\cdot \vec{\mu}_j}{r_{ij}^3}-\frac{3 (\vec{r}_{ij}\cdot \vec{\mu}_i)(\vec{r}_{ij}\cdot \vec{\mu}_j)}{r_{ij}^5}\right)\nonumber\\
 &\mbox{with} &i\ne j,\, i= j\ne\pm1
\end{eqnarray}
 where $\vec{\mu}_i$ is the dipole moment and $r$ is the distance between the interacting dipoles with constants: $\epsilon=13059$ cm$^{-1}$ and $C=519\,044\,  \AA^3$ cm$^{-1}$. The LH1 complex has a spectrum maximum near 875 nm, with inter-complex distances and nearest neighbor interactions equal to the ones provided in the B850 LH2 ring. The molecular nature of chromophores involve vibrational degrees of freedom that provide a manifold within each electronic states which should be accounted. However, given these parameters within a harvesting complex,  intra-complex energy transfer involves sub-picosecond time-scales, that imply relevance of electronic quantum features over the influence of thermalization in the vibrational manifold.  If  decoherence sources are not important within the  time-scale of excitation dynamics on donor or acceptor aggregates, they can be  known from the eigenvector problem $H\vert \phi_{k}\rangle=E_{k}\vert \phi_{k}\rangle$ whose solution provides excitonic fully delocalized states
\begin{equation}\label{state}
\vert{\phi_k}\rangle=\sum_n a_{k,i}\vert i\rangle
\end{equation}
 considered to best describe the B850 ring. A slightly less interaction strength allows thermalization through the vibrational manifold in comparable time-scales, degrading delocalization in the B800 ring  over 2-3 pigments, however,  able to improve robustness of B800$\rightarrow$B850 energy transfer \citep{silbey2004}.

\begin{figure}
\centering
\includegraphics[width=7 cm,height=5 cm]{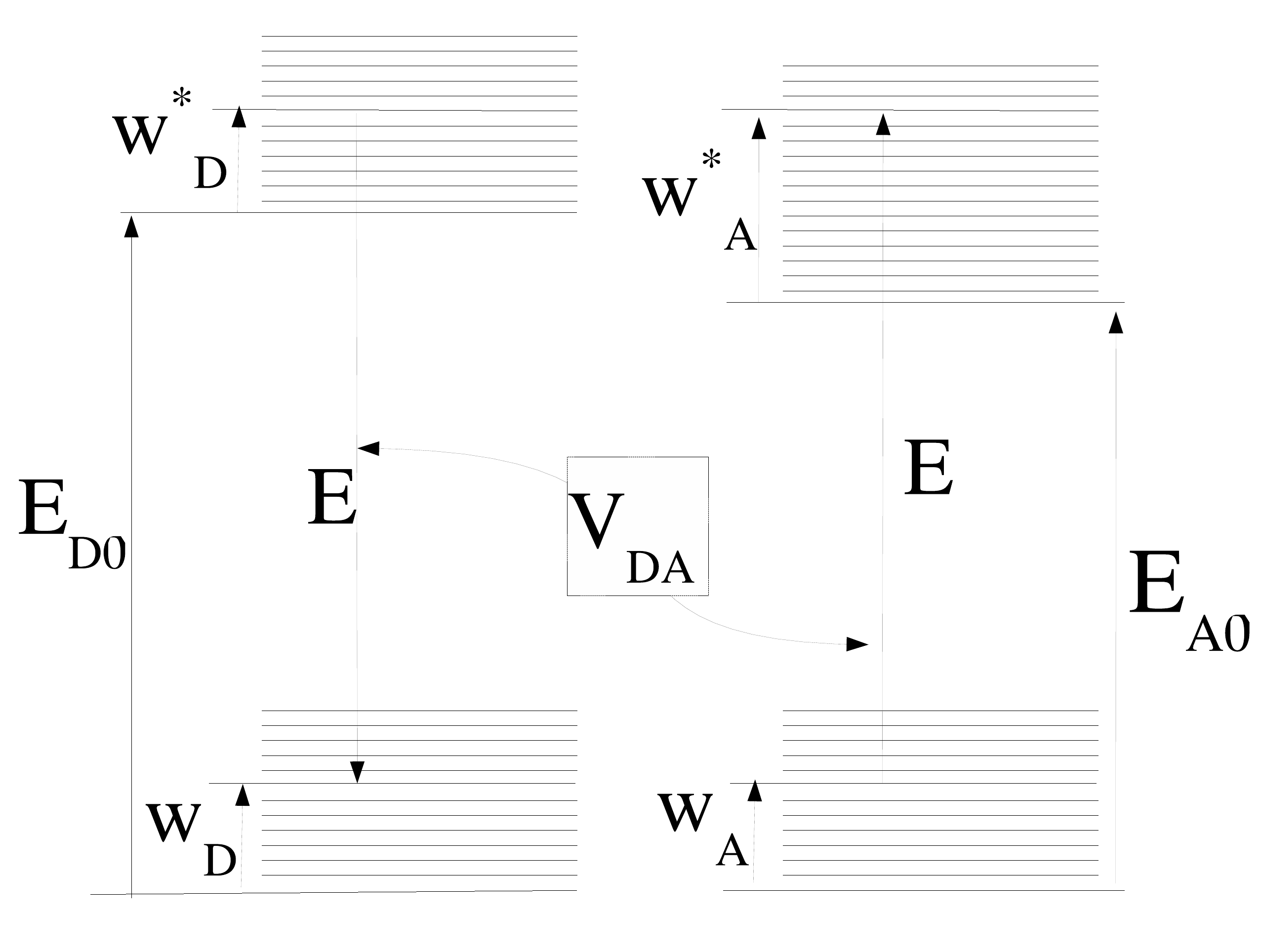}
\caption{Energy level scheme of donor and acceptor molecules. Although the zero phonon line might be different between both, energy conservation on the transfer applies due to vibrational levels.}\label{fig5}
\end{figure}

The fact that states involved in excitation transfer in molecules include the vibrational manifold,  makes enormous the Hilbert space over which the sums of eq.(\ref{vij}) should be  formally performed. When  the transfer occurs with chromophores that belong to different complexes, say inter-complex energy transfer, rapid thermalization out stands direct Coulomb mechanism, and no defined phase relationship between donor and acceptor electronic states is expected. The hermitian nature of excitonic exchange is replaced due to decoherence, by a rate to describe   electronic excitation transfer  within thermalization in the vibrational manifold. According to Fermi's golden rule, adapted to the vibrational continuum, the rate of transfer is given by \citep{review}:
\begin{eqnarray}
k_{DA}&=&\frac{2\pi}{\hbar}\int dE\int_{E_{A_0}} dw_A\int_{E_{D^*_0}} dw^*_D  \frac{g_D^*(w^*_D)\exp(-w_D^*/k_BT)}{Z_D^*}\frac{g_A(w_A)\exp(-w_A/k_BT)}{Z_A}\vert\tilde{U}_{DA}\vert^2\nonumber\\
 & &\label {jda}
\end{eqnarray}
interpreted as a the sum  of Coulomb contributions from electrons at donnor (D) and acceptor (A) aggregates in ground or excited states (these latter dennoted by $^*$), $\tilde{U}_{DA}=\langle \Psi_{D^*}\Psi_A\vert V_{DA}\vert \Psi_D\Psi_{A^*}\rangle$, weighted by Boltzmann factor   ($k_B$ is Botlzmann constant) and vibrational manifold multiplicity ($g_{D(A)}(w_{D(A)})$ for donnor (acceptor)) at the electronic energies $w_{D(A)}$ measured from the donnor (acceptor) zero phonon lines $E_{D(A)_0} $.  With the Born-Oppenheimer approximation \citep{fcondon,review}  $\ket{\Psi}$ is assumed as products of electronic $\ket{\phi}$ and vibrational $\ket{\chi}$ molecular states:
\begin{eqnarray} \label{hatUda}
\hat{U}_{DA}&\approx& \langle \phi_{D^*}\phi_A\vert V_{DA}\vert \phi_D\phi_{A^*}\rangle
\times\langle \chi(w_D^*)\vert\chi(w_D)\rangle \langle \chi(w_A)\vert\chi(w_A^*)\rangle\nonumber\\
 &\approx & U_{DA}\langle \chi(w_D^*)\vert\chi(w_D)\rangle \langle \chi(w_A)\vert\chi(w_A^*)\rangle\label{hatUda}
\end{eqnarray}
where  $U_{DA}= \langle \phi_{D^*}\psi_A\vert V_{DA}\vert \psi_D\psi_{A^*}\rangle.$ Using the above approximation, the expression (\ref{jda}) now including the overlap between vibrational levels can be cast in a  more illustrative form:
\begin{equation}\label{kda}
k_{DA}=\frac{2\pi}{\hbar}\vert U_{DA}\vert^2 \int dE\,G_D(E)G_A(E)
\end{equation}
Here, $G_{D}(E)$ and $G_A(E)$ are often called the Franck-Condon weighted and thermally averaged combined density of states. Explicitly:
\begin{eqnarray}
G_D(E)&=&\int_{E_{D_0^*}}dw_D^* \frac{g_D^*(w_D^*)\exp(-w_D^*/k_BT)\vert\langle \chi(w_D^*)\vert\chi(w_D)\rangle\vert^2}{Z_D^*}\nonumber
\end{eqnarray}
with equal expression for acceptor molecule by replacement $D\rightarrow A$.  F\"orster showed \citep{forster1965} that these distributions are related to extinction coefficient $\epsilon(E)$ and fluorescence spectrum $f_D(E)$ of direct experimental verification:
\begin{eqnarray}
\epsilon(E)=\frac{2\pi N_0}{3 \ln 10\hbar^2\,nc}\vert \mu_A\vert^2 E\, G_A(E),\quad f_D(E)=\frac{3\hbar^4c^3\tau_0}{4n}\vert \mu_D\vert^2E^3 G_D(E)
\end{eqnarray}
where $N_0=6.022\times 10^{20}$ is the number of molecules per mol per cm$^3$, $n$ is the refractive index of the molecule sample, $c$ the speed of light, and $\tau_0$ the mean fluorescence time of the donnor excited state. For normalized spectra
\begin{equation}
\hat{\epsilon_A}(E)=\frac{\epsilon_A(E)}{\int dE\,\epsilon_A(E)/E},\quad \hat{f_D}(E)=\frac{f_D(E)}{\int dE\,f_D(E)/E^3}
\end{equation}
and from the relations  $w_D=E_{D_{0^*}}+w_D^*-E$ and $w_A^*=-E_{A_0}+w_A+E$ schematically presented in Fig.\ref{fig5}, the F\"orster rate is cast
\begin{equation}
k_{DA}=\frac{2\pi}{\hbar}\vert U_{DA}\vert^2 \int dE \frac{\hat{\epsilon}_A(E)\hat{f}_D(E)}{E^4}.
\end{equation}
Therefore, whenever fluorescence and absorption spectra are available, an estimate for the  excitation transfer rate can be calculated.

Thermalization occurs firstly in the vibrational manifold of the electronic states involved. Due to the greater energy gap of electronic transitions compared with the one of vibrational nature,   on a longer time-scale thermalization also occurs in excitonic states. Accordingly, the calculation  in eq.(\ref{hatUda}) involves an statistical thermal  mixture, explicitly:
\begin{equation}
\rho=\frac{1}{\Tr{\cdot}}\sum_{k}\exp(-E_{k}/k_BT)\vert \phi_k\rangle\langle \phi_k\vert
\end{equation}
where $\Tr{\cdot}$ is trace of the numerator operator, used to normalize the state. Hence, in a straightforward fashion
\begin{eqnarray}\label{Uda}
U_{DA}=\mbox{Tr}\{\rho V_{DA}\}&=&\frac{1}{\sum_k\exp(-E_k/k_BT)}\sum_{k}\sum_{p}\exp(-E_k/k_BT)\langle \phi_k\vert V_{DA}\vert\phi_p\rangle\nonumber\\
 &= &\frac{1}{\sum_k\exp(-E_k/k_BT)} \sum_{k,p,i,j}a_{k,i}a^*_{p,j}V_{ij}\label{Uda}
\end{eqnarray}
where the element $\langle\phi_k\vert V_{DA}\vert\phi_p\rangle$ are the elements of interaction among exciton states in molecules on different complexes and use is made of the individual contributions of excitonic states in chromophore's site basis  eq.(\ref{state}).

Summarizing, excitation transfer occurs through induced dipole transfer, among BChls  transitions.  The common inter-complex BChl distances 20-100 $\AA$ \citep{scheuring2005,bahatyrova2004} cause excitation transfer to arise through the Coulomb interaction on the picosecond time-scale \citep{review}, while vibrational dephasing destroys coherences within a few hundred femtoseconds \citep{Engel2010,lee2007}. As noted, the Coulomb interaction as dephasing occurs, makes the donor and acceptor phase become uncorrelated pointing into a classical rate behavior.  Transfer rate measures from pump-probe experiments agree with the just outlined generalized F\"orster calculated rates \citep{review}, assuming intra-complex delocalization along thermodynamical equilibrium. LH2$\rightarrow$LH2 transfer has not been measured experimentally, although an estimate of $t_{22}=10$ ps has been calculated \citep{review}. LH2$\rightarrow$ LH1 transfer has been measured for {\it R. Sphaeroides} as $t_{21}= 3.3$ps \citep{hess1995}. Due to formation of excitonic states \citep{silbeyreview}, back-transfer LH1$\rightarrow$ LH2 is enhanced as compared to the canonical equilibrium rate for a two-level system, up to a value of $t_{12}=15.5 $ps. The LH1$\rightarrow$LH1 mean transfer time $t_{11}$ has not been measured, but the just mentioned generalized F\"orster calculation \citep{ritz2001} has reported an estimated mean time $t_{11}$ of 20 ps.  LH1$\rightarrow$ RC transfer occurs due to ring symmetry breaking through optically forbidden (within ring symmetry) second and third lowest exciton lying states \citep{hu1997}, as suggested by agreement with the
experimental transfer time of 35-37 ps  at 77 K \citep{visscher1989,visscher91}.  Increased spectral overlap at room
temperature improves the transfer time to $t_{1,RC}=25$ ps  \citep{vGrondelle1994}. A photo-protective design makes the
back-transfer from an RC's fully populated lowest exciton state to higher-lying LH1 states occur in a calculated time of
$t_{RC,1}=$8.1 ps \citep{hu1997}, close to the experimentally measured 7-9 ps  estimated from decay kinetics after RC excitation
\citep{timpmann1993}.

Table \ref{times} shows the results of mean transfer times presented in Ref.\citep{ritz2001}  through the above mentioned calculation, compared with the experimental evidence  restricted to different complex kind from the spectral resolution requirement of pump-probe spectroscopy. Since LH1$\leftrightarrow$ LH1 and LH2$\leftrightarrow$ LH2 transfer steps involve equal energy transitions, no experimental evidence is available regarding the rate at which these transitions occur. The experimentally determined B800$\rightarrow$ B850 rate was 1/700fs \citep{shreve1991}. The  inter-complex transfer rate between LH2$\rightarrow$ LH1 have been determined experimentally to be 1/3.3ps \citep{hess1995}. Experimentally, LH1$\leftrightarrow$ RC forward transfer rate ranges between   1/50ps and 1/35ps, while back-transfer rate ranges between 1/12ps and 1/8ps \citep{visscher1989,timpmann1993,timpmann1995}.  It is interesting to note that exists a two fold difference in the experimental and theoretical determined LH2$\rightarrow$LH1, ascribed to BChl{\it a} Q$_y$ dipole moment underestimation. It is assumed for theoretical calculation a value of 6.3 Debye, while a greater BChl{\it a} Q$_y$ dipole moment in Prostecochloris aestuarii (not  a purple bacterium) of 7.7 Debye has been determined \citep{ritz2001}. On the other hand, LH1$\rightarrow$RC theoretical calculation  gives a greater value for tranfer rate, thought to arise due to an overestimate of LH1 exciton delocalization \citep{ritz2001}. This rate decreases when delocalization is assumed over fewer BChl's, therefore, further research is needed to understand the effect of decoherence sources (static inhomogeneities and dynamical disorder due to thermal fluctuations) on the delocalization length.

\begin{table}[h]
\centering
\begin{tabular}{|c|c|c|c|}
\hline
from$\backslash$ to & LH1 & LH2 & RC\\
\hline
LH1 &20.0/N.A. &15.5/N.A.  &15.8/30-50  \\
\hline
LH2 &7.7/3.3&10.0/N.A. & N.A\\
\hline
RC &8.1/8 & N.A & N.A \\
\hline
\end{tabular}
\caption{Theoretical estimation/experimental evidence of inter-complex transfer times in picoseconds. N.A are not available data.}\label{times}
\end{table}

\subsection*{1.3 Dissipation}
Excitation in chromophores might be dissipated by two main mechanisms. The first is fluorescence, where the electronic excited state has a finite lifetime on the nanosecond time-scale, due to its interaction with the electromagnetic vacuum \citep{scully}. The second is internal conversion, where the electronic energy is transferred to vibrational degrees of freedom.
\begin{figure}
\vspace{0 cm}
\hspace{0 cm}
\centering
\includegraphics[width=12 cm]{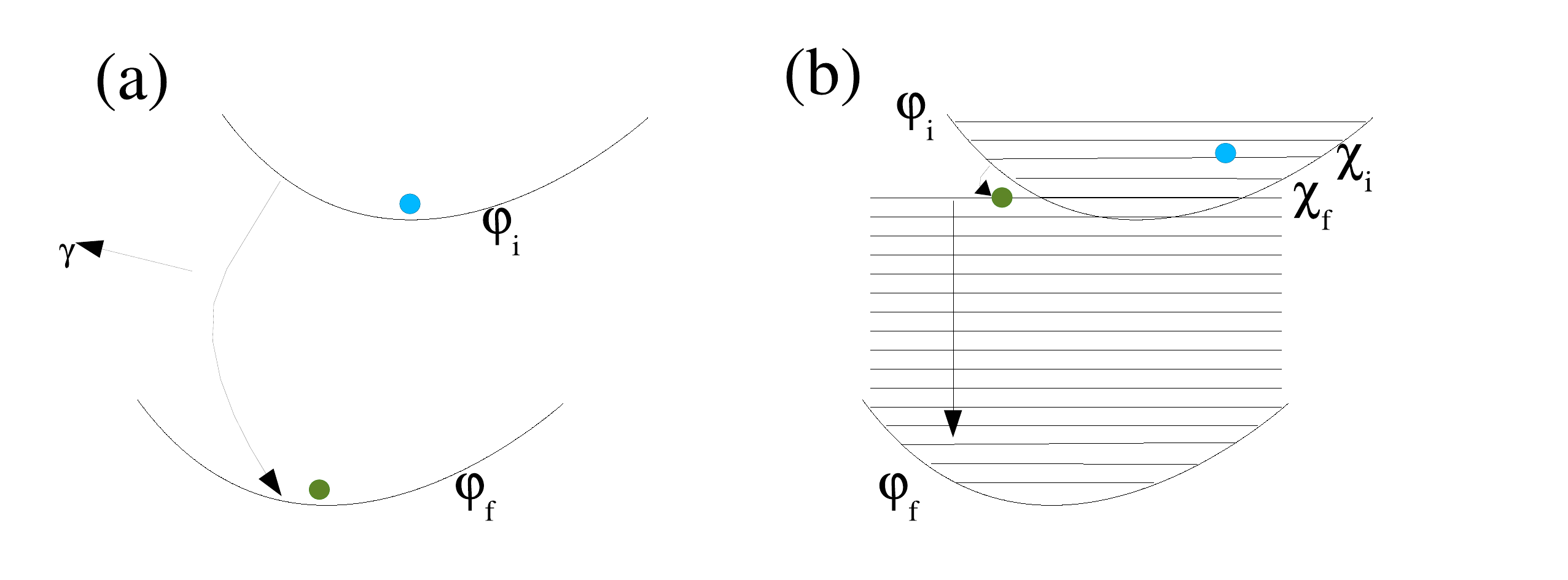}
\caption{Dissipation mechanisms. In (a), the electron de-excites due to its interaction with the quantized electromagnetic vacuum field through a fluorescent photon. In (b) internal conversion mechanism, where the vibrational levels overlap induces a transition between electronic excited and ground states. Dissipation overcomes when thermal equilibrium is reached in the vibrational manifold of electronic ground state.}\label{dissplot}
\end{figure}

Within the Born-Oppenheimer approximation, the molecular state $\Psi$, can be decomposed into purely electronic $\phi$ and (nuclear) vibrational $\chi$ states. The transition probability between initial state $\Psi_i$ and final state $\Psi_f$, is proportional to $\langle \Psi_i \vert H\vert \Psi_f\rangle\propto \langle\chi_i\vert\chi_f\rangle$. Note that $\chi_i$ ($\chi_f$) are vibrational levels in the ground (excited) electronic state manifold (see Fig.\ref{dissplot}). If the energy difference is small, and the overlap between vibrational levels of different electronic states is appreciable, the excitation can be transferred from the excited electronic state, to an excited vibrational level in the ground electronic state. This overlap increases with decreasing energy difference between electronic states. As  higher electronic levels have smaller energy difference among their zero phonon lines,  internal conversion process is more probable the higher energy electronic states have.  Fluorescence and internal conversion between first excited singlet and ground electronic states,  induce dissipation in a range of hundreds of picoseconds and a few nanoseconds. Numerical simulations are performed with a dissipation time including both fluorescence and internal conversion of $1/\gamma_D= 1$ ns, also used in \citep{ritz2001}.

\subsection*{1.4 Special pair (SP) ionization}\label{Pion}

From  the LH1 complex excitation  reaches the RC, specifically the special pair (SP) dimer.   The excitation can   be transferred back to its surrounding LH1, or initiate a chain of ionizations along the A branch, probably, due to a tyrosine residue strategically positioned instead of a phenylalanine present in the B branch \citep{norrisflemming85}. Once the special pair is excited, it has been determined experimentally \citep{flemming1988} that takes 3-4 ps for the special pair to ionize and produce a reduced bacteriopheophytin, $H_A^-$, in a reaction $SP^*\rightarrow SP^+H_A^-$. This reaction initiates an electron hop, to a quinone $Q_A$ in about 200 ps, and to a second quinone, $Q_B$ if available. Initially, the ionized quinol $Q_B^+$ captures an introcytoplasmic proton and produces hydroxiquinol $Q_BH$, which after a second ionization that produces $Q_BH^+$  to form quinol $Q_BH_2$. After any SP ionization a neutrality restablishment is required, provided by the cytochrome $cyt$ charge carrier. After SP ionization, the cytochrome diffuses from the bc1 complex to a RC in order to replenish its neutrality SP$^+\rightarrow$SP, within several microseconds \citep{milano2003}.  The first electron transfer step $P^*\rightarrow P^+$ occurs in the RC within $t_{+}=$3 ps, used for quinol
($Q_BH_2$) production \citep{review}.

\subsection*{1.5 Quinone-quinol cycling}\label{RCcycle}

The RC cycling dynamics also involves undocking of $Q_BH_2$ from the RC due to  lower affinity among RC and this new product.  Quinol  starts a migration to the $bc1$ complex where enables the ionization of  the cytochrome $cyt$ charge carrier, while a new quinone $Q_B$ molecule  docks into the RC.  The time before  quinol unbinds, and a new $Q_B$ is available, has been reported within  milliseconds \citep{savoti1997} to highlight quinol removal o as the rate limiting step \citep{savoti1997} if compared to special pair restablishment.

Even though  it has been reported that excitation dynamics change as a function of the RCs state \citep{borisov1985, comayras2005}, at a first glance the several orders of magnitude difference among the picosecond transfer, the nanosecond dissipation and the millisecond RC cycling, seems to disregard important effects due to these mechanisms' interplay.  However, the quinol-quinone dynamics leaves the RC unable to promote further quinol production and eventually enhances the influence of dissipation of a wandering excitation, evident when none RC is available and the unique fate of any excitation is to be dissipated.

Interestingly, the quinone-quinol mechanism has been well established and thought to be of priority on adaptations of bacteria, that seem to respond to its dynamics. For instance, an observed trend for membranes to form clusters of same complex type  \citep{scheuring2004} seems to affect diffusion of quinones, enhanced when, due to higher mobility of LH1s, left void spaces help travel quinones to the periplasm. Negligible mobility of LH2s in their domains, would restrict metabolically active quinones to LH1 domains \citep{sturg2006}. Easier diffusion of quinones, quinol and cytochromes promotes higher availability of charge carriers in RC domains under LLI conditions, increasing the rate at which RCs can cycle.  The RC cycling dynamics and its connection to the membranes performance has been accounted in \citep{PRLCaycedo-Soler,NJPCaycedo-Soler} in a quantitative calculation to understand the effect of core clustering and stoichiometry variation in the RC supply or in the efficiency of the membranes from experimentally obtained Atomic Force Microscopy images, to be presented in this chapter.

\section*{2. Exciton kinetics}\label{model}
\begin{figure}
\centering
\includegraphics[width=0.7\columnwidth]{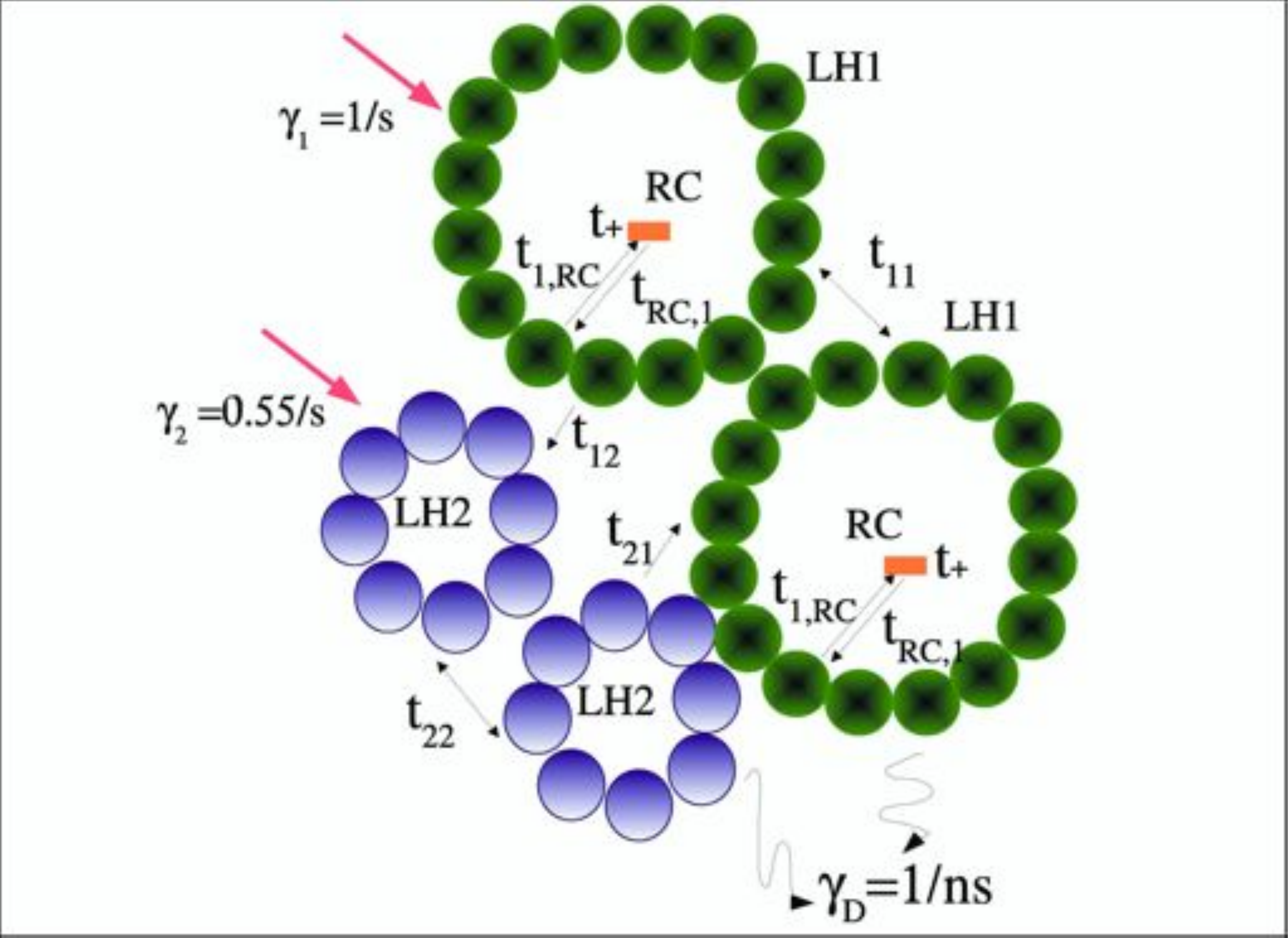}
\vspace{-0.01 cm} \caption{Schematic of the biomolecular
photosynthetic machinery in purple bacteria, together with
relevant inter-complex mean transfer times $t_{ij}$, dissipation
rate $\gamma_D$, and normalized light intensity rate
$\gamma_{1(2)}$}\label{rates}
\end{figure}

Figure \ref{rates} summarizes the relevant biomolecular complexes in purple bacteria {\it Rsp. Photometricum} \citep{scheuring2004}, together with experimental-- theoretical if the former are not available-- timescales governing the excitation kinetics: absorption and transfer; and reaction center dynamics: quinol removal.

\subsection*{2.1 Model}

The theoretical framework  used to describe the excitation
transfer must be built around the experimental (if available) and theoretical parameters just
outlined.  Remind that the thermalization process occurs faster than inter-complex energy transfer, and provides the support to rely in a classical hopping process, since  phase information is lost well within the time frame implied by direct Coulomb coupling. Accordingly, we base our analysis on a classical random walk for excitation dynamics along the full vesicle, by considering a collective state with $N=N_2+2N_1$ sites -- resulting
from $N_2$ LH2s, $N_1$ LH1s and hence $N_1$ RC complexes in the vesicle. The state vector $\vec{\rho}=(\rho_1,\rho_2,...,\rho_{M})$ has in each element  the probability of occupation  of a collective state comprising several excitations. If a single excitation is allowed in each complex, both excited and ground states of any complex should be accounted and the state space size is $M=\underbrace{2\times2\times2...}_N=2^N$. On the other hand, if only one excitation that wanders in the whole network of complexes is allowed, a site basis can be used where each element of the state vector gives the probability of residence in the respective complex, and reduces the state vector size  to $M=N$. In either case the state  vector time evolution
obeys a master equation
\begin{equation}\label{meq}
\partial_t\rho_i(t)=\sum_{j=1}^M G_{i,j}\rho_j(t).
\end{equation}
where $G_{i,j}$ is the transition rate from a collective state or site $i$-- whether many  or a single excitation are accounted, respectively -- to another collective state or site $j$.  Since the transfer rates do not depend on time, this yields
a formal solution $\vec{\rho}(t)=\tilde{e^{G t}}\vec{\rho}(0)$. However, the required framework depends on exciton abundance  within the whole chromatophore at the regime of interest.

For instance, purple bacteria ecosystem concerns several meters depths, and should be reminded as a  low light intensity environment. Within a  typical range of 10-100 W/m$^2$ and a commonly sized chromatophore having $\approx 400$ LH complexes, eq.(\ref{gammaa}) leads to an absorption rate $\gamma_A\approx$100-1000 s$^{-1}$, which compared with  the dissipation mechanisms (rates of $\approx 10^9$ s$^{-1}$) imply that an absorption event  occurs and then the excitation will be trapped by a RC or become dissipated within a nanosecond, and other excitation will visit the membrane not before some milliseconds have elapsed.  However, it is important to remind the nature of thermal light where the possibility of having bunched small or long inter-photon times is greater than evenly spread, with greater  deviations from poissonian statistics the grater its mean intensity is. Therefore, regardless of such deviations, under the biological light intensity conditions, the event of two excitations present simultaneously along the membrane will rarely occur and a single excitation model is  accurate.

\subsection*{2.2 Small architectures}
Small absorption rates lead to single excitation dynamics in the
whole membrane, reducing the size of $\vec{\rho}(t)$ to the total
number of sites $N$. The probability to have one excitation at a
given complex initially, is proportional to its absorption cross
section, and can be written as
$\vec{\rho}(0)=\frac{1}{\gamma_A}(\underbrace{\gamma_1,...}_{N_1},\underbrace{\gamma_2,...}_{N_2},\underbrace{0,..}_{N_1})$,
where subsets correspond to the $N_1$ LH1s, the $N_2$ LH2s and the
$N_1$ RCs respectively.

\subsubsection*{2.2.1 Complexes arrangement: architecture}
To gain physical insight on the global behavior of the harvesting membrane, our interest lies in the probability to have an excitation at a  given complex kind $k\,\in$ LH1,LH2 or RC, namely  $\hat{p}_k$, given that at least one excitation resides in the network:
\begin{equation}
\hat{p}_{k}(t)=\frac{\rho_{k}(t)}{\sum_{i=1}^{N} \rho_{i}(t)}\ \ .
\end{equation}

\begin{figure}
\centering
\includegraphics[width=0.3\columnwidth]{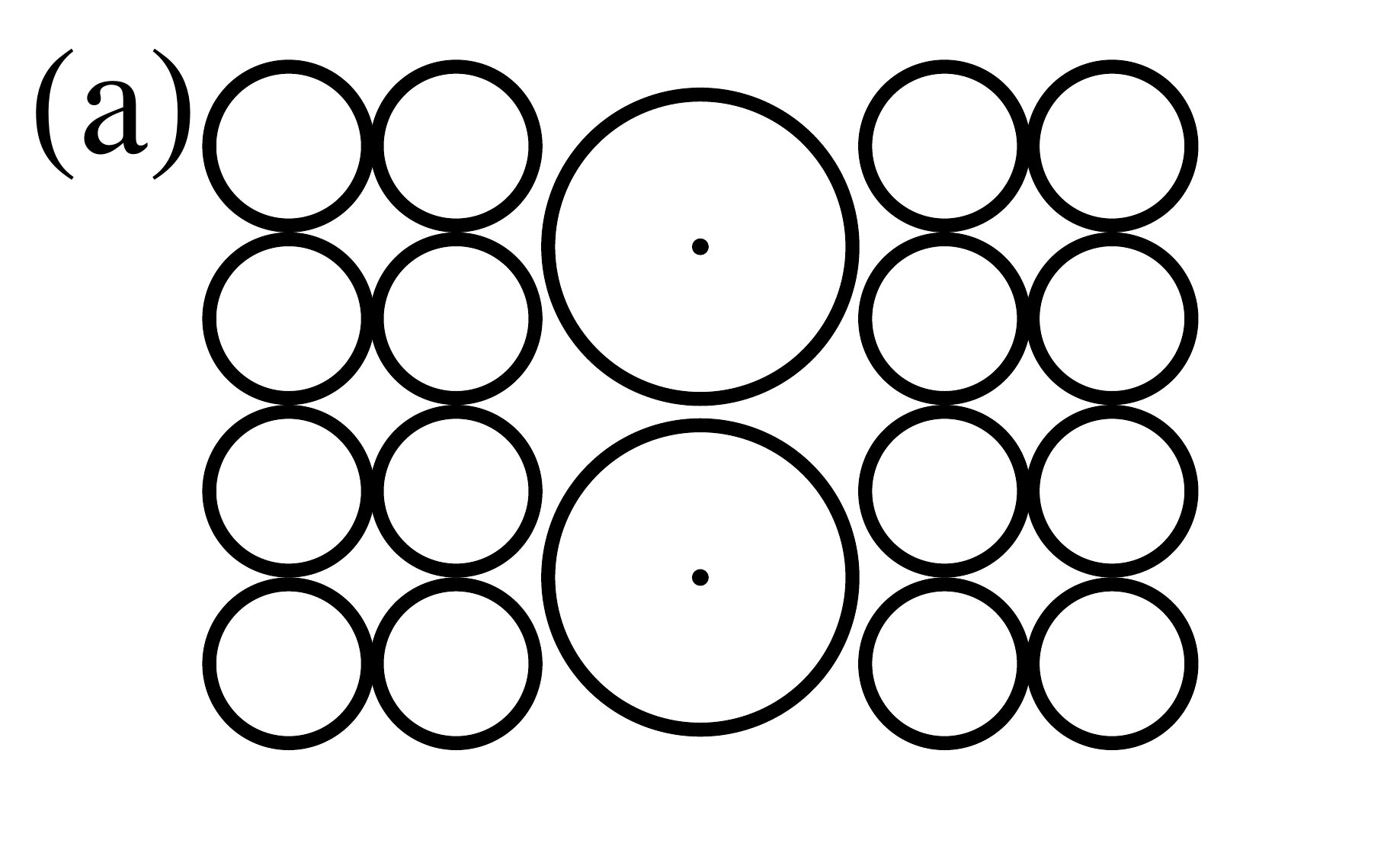}
\includegraphics[width=0.3\columnwidth]{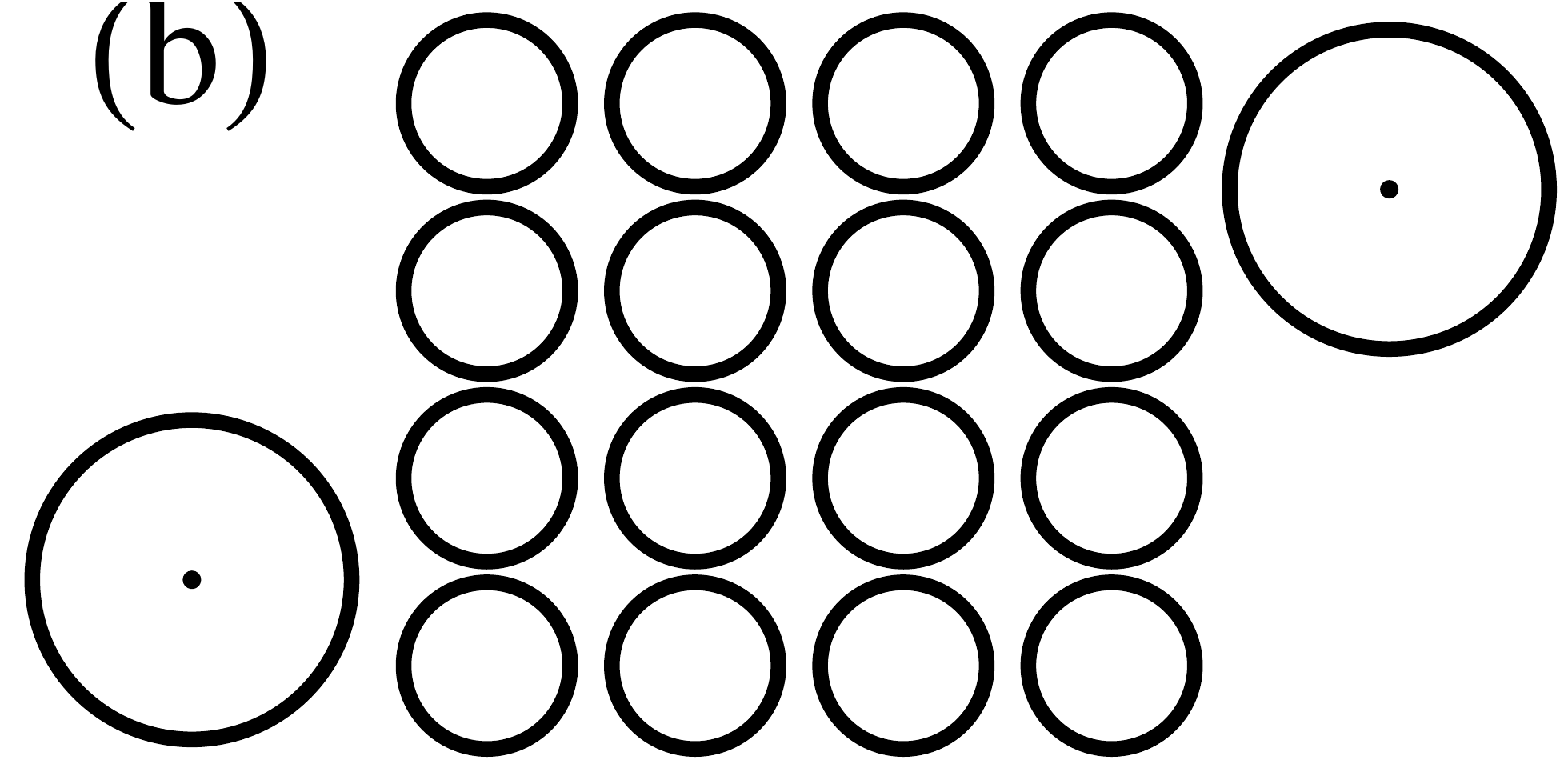}
\includegraphics[width=0.3\columnwidth]{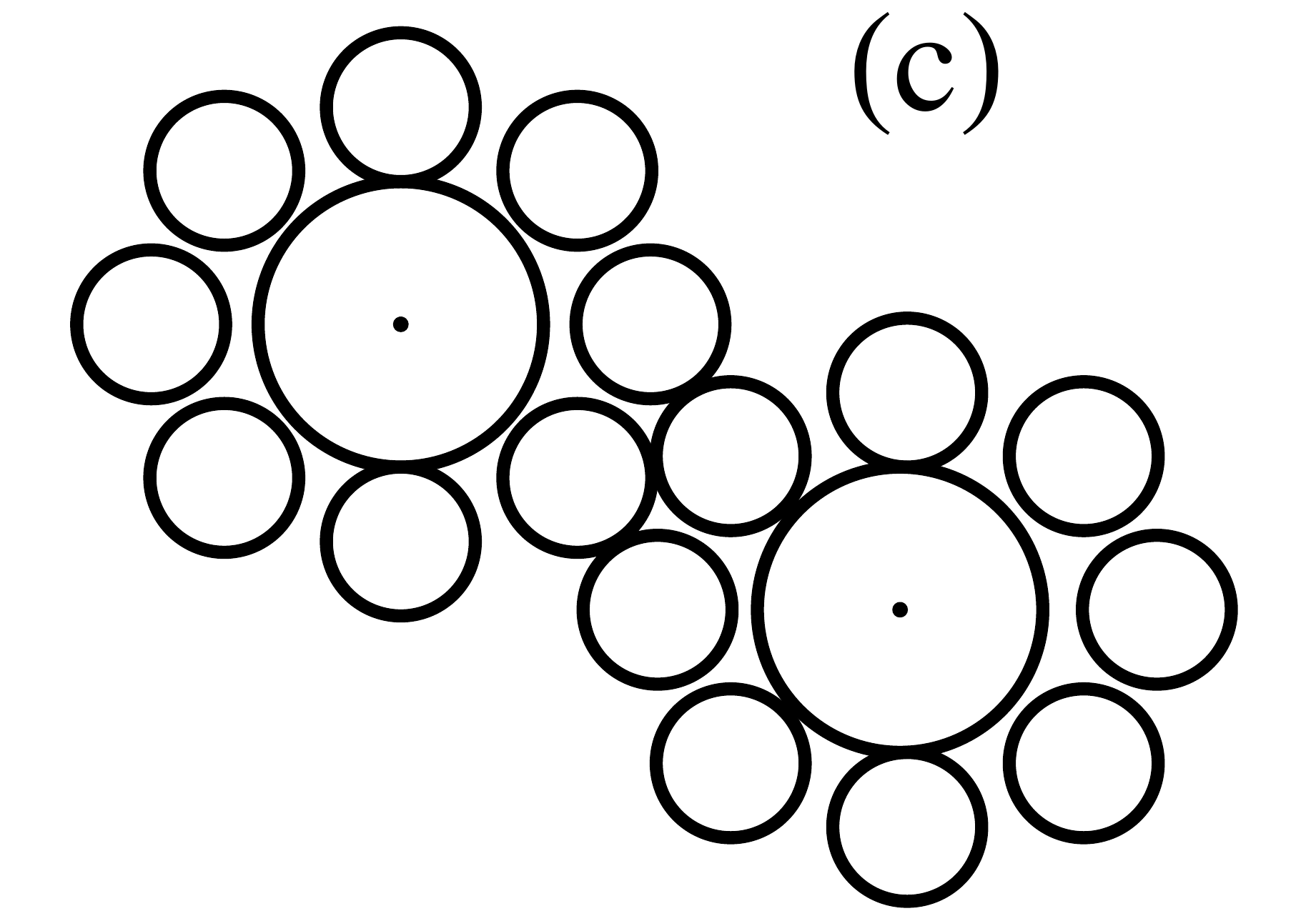}\\
\begin{minipage}{0.65\columnwidth}
\hspace{-2 cm}
\includegraphics[width=1\columnwidth]{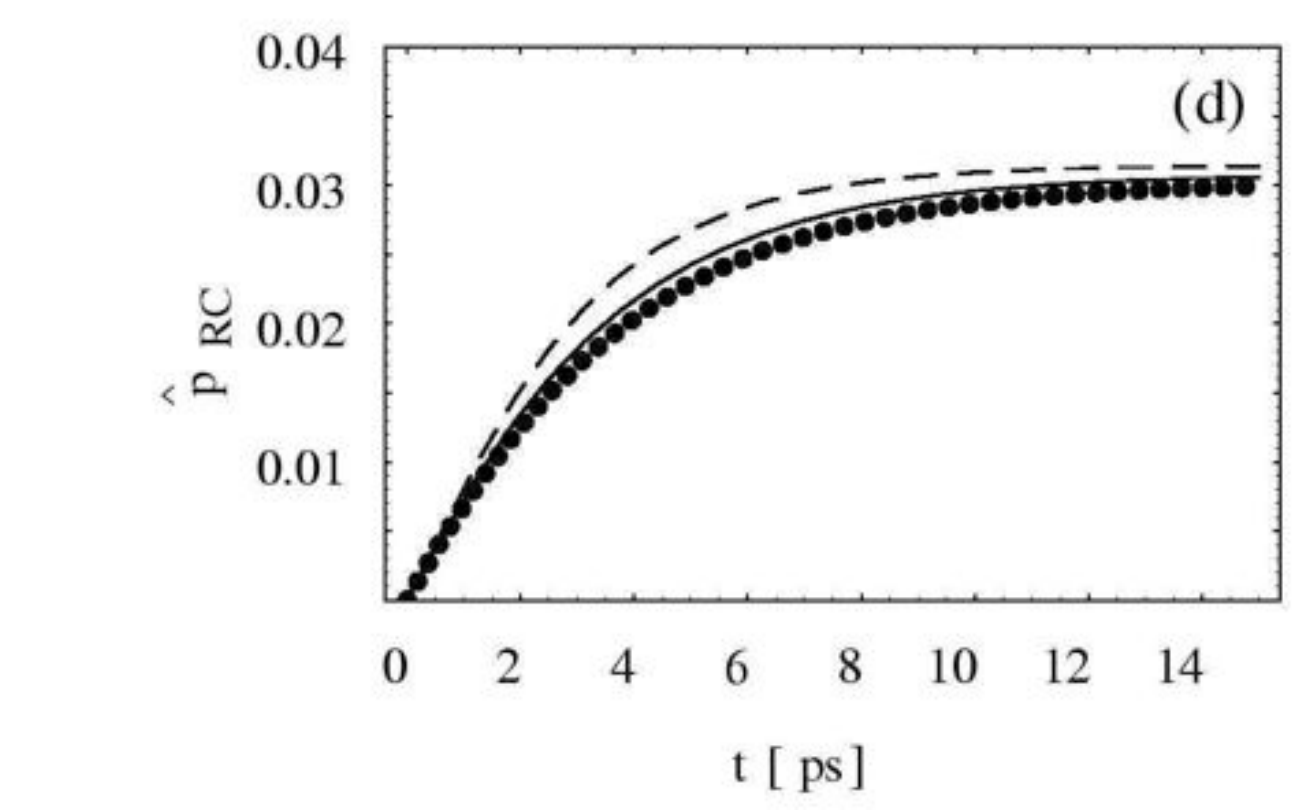}
\end{minipage}\hspace{-2 cm}
\begin{minipage}{0.31\columnwidth}
\includegraphics[width=1\columnwidth]{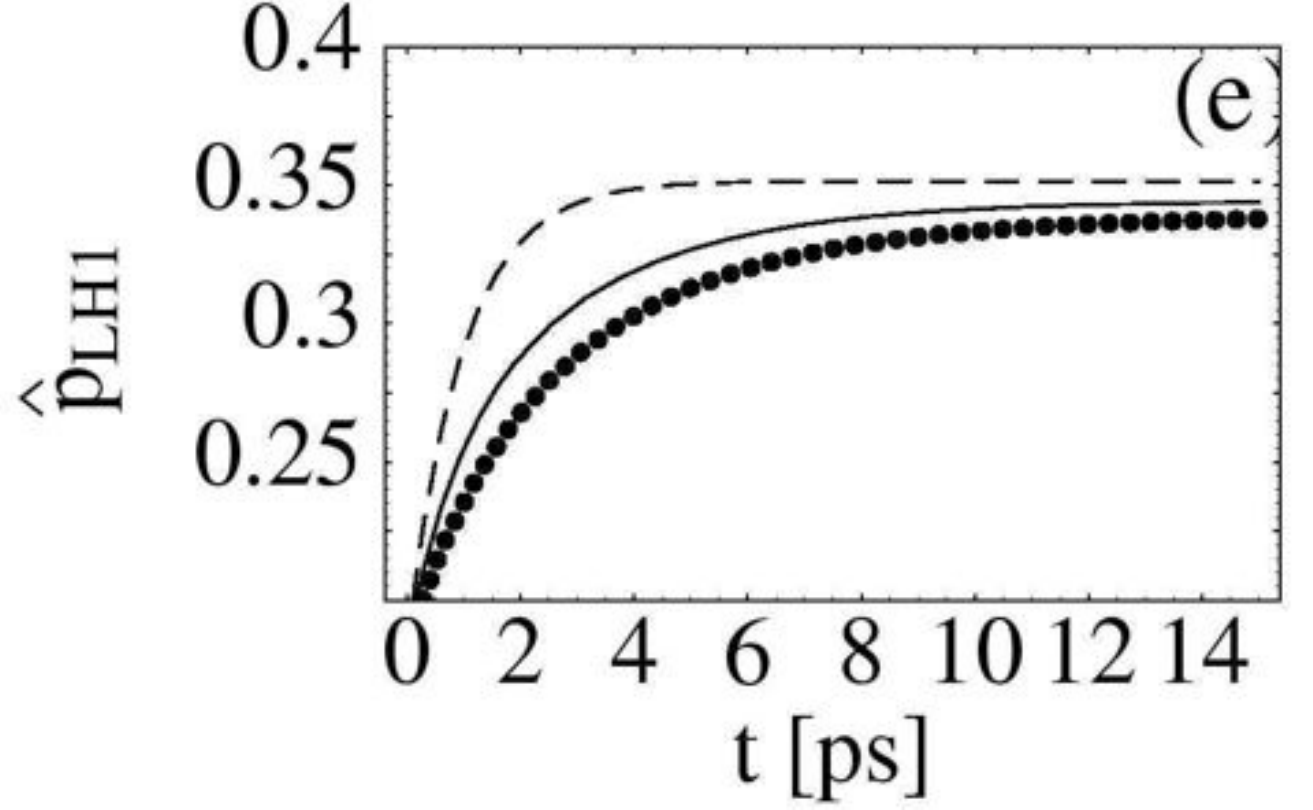}
\includegraphics[width=1\columnwidth]{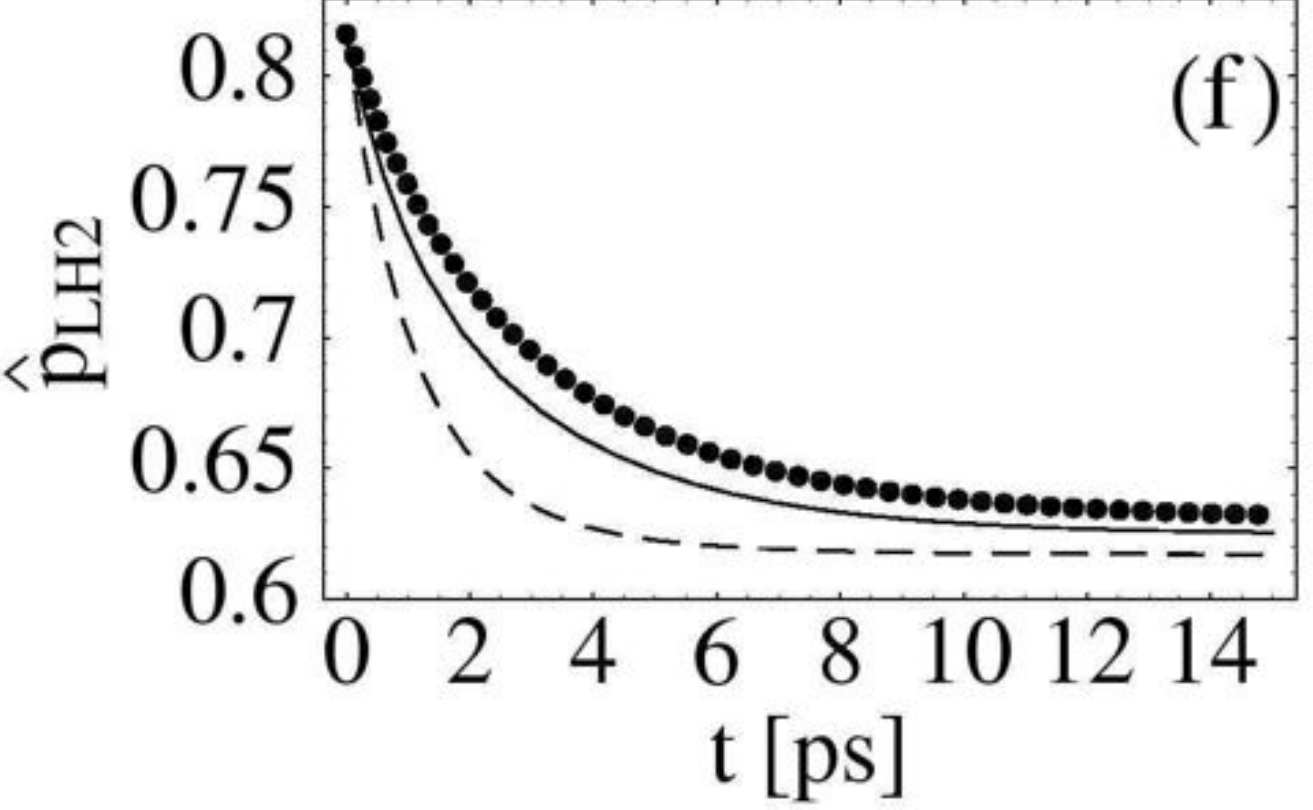}
\end{minipage}
\caption{Top panel: Three example small network architectures. The bottom panel shows the normalized probabilities for finding an
excitation at an RC (see (d)), an LH1 (see (e)), or an LH2 (see (f)). In panels (d)-(f), we represent
these architectures as follows: (a) is a continuous line; (b) is a dotted line; (c) is a dashed line.}\label{archs0}
\end{figure}

The effects that network architecture might have on the model's dynamics, are studied with different arrangements of complexes in
small model networks, focusing on architectures which have
the same amount of LH1, LH2 and RCs as shown in the top panel of
Fig.\ref{archs0}(a), (b) and (c). The bottom panel
Fig.\ref{archs0} (d)-(e)-(f) shows that  $\hat{p}_{k}$ values for RC,
LH1 and LH2 complexes, respectively. First, it is important to notice that excitations trend is to stay within LH1 complexes, and not in the RC. Fig.\ref{archs0}(d) shows that
the highest RC population is obtained in configuration (c),
followed by configuration (a)  and (b) whose ordering relies in
the connectedness of LH1s to antenna complexes. Clustering of LH1s
will limit the number of links to LH2 complexes, and reduce the
probability of RC ionization. For completeness, the probability of
occupation in LH1 and LH2 complexes  (Figs.\ref{archs0}(e) and (f),
respectively), shows that increased RC occupation benefits from population imbalance between
LH1 enhancement and LH2 reduction. As
connections among antenna complexes become more favored, the probability
of finding an excitation on antenna complexes will become smaller,
while the probability of finding excitations in RCs is
enhanced. This preliminary result, illustrates that if the apparent requirement   to funnel excitations down to RCs in bacterium  were of primary importance, the greatest connectedness of LH1-LH2 complexes should occur in nature as a consequence of millions of years evolution. However, as will be presented, the real trend to form LH1 clusters, reduces its connectedness to antenna LH2 complexes and somehow pinpoints other mechanisms as the rulers of harvesting membranes conformation and architecture.

\subsubsection*{2.2.2 Relative amount of complexes: Stoichiometry}

\begin{figure}
\centering
\includegraphics[width=10 cm]{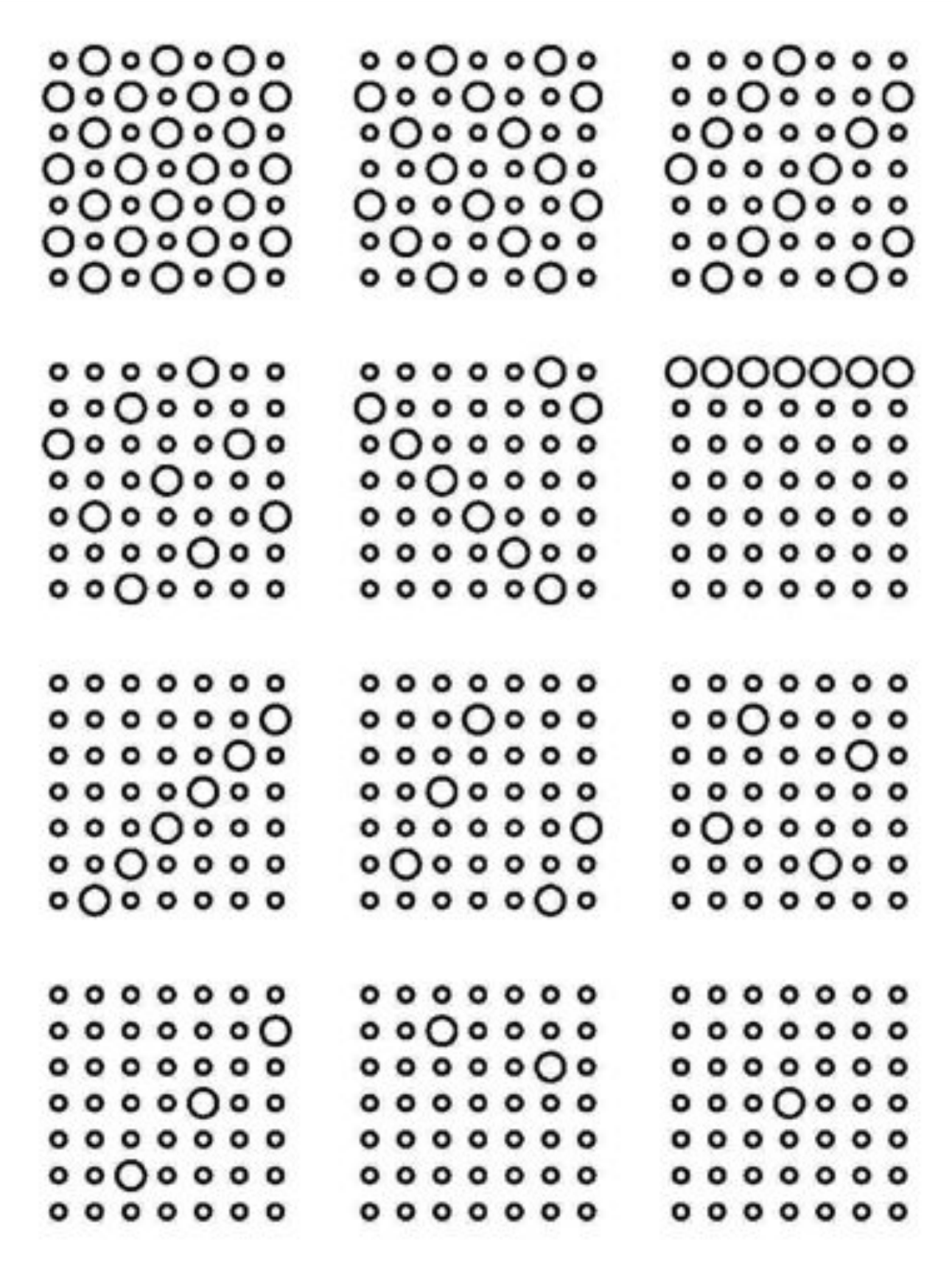}
\caption{Networks with different stoichiometries, from left to right, top to bottom, $s$=$\{$1.04, 2.06, 3.08, 4.44, 5.125, 6, 7.16, 8.8, 11.25, 15.33, 23.5, 48$\}$, and equal number  of harvesting complexes.}\label{archs2}
\end{figure}

\begin{figure}
\centering
\includegraphics[width=6. cm]{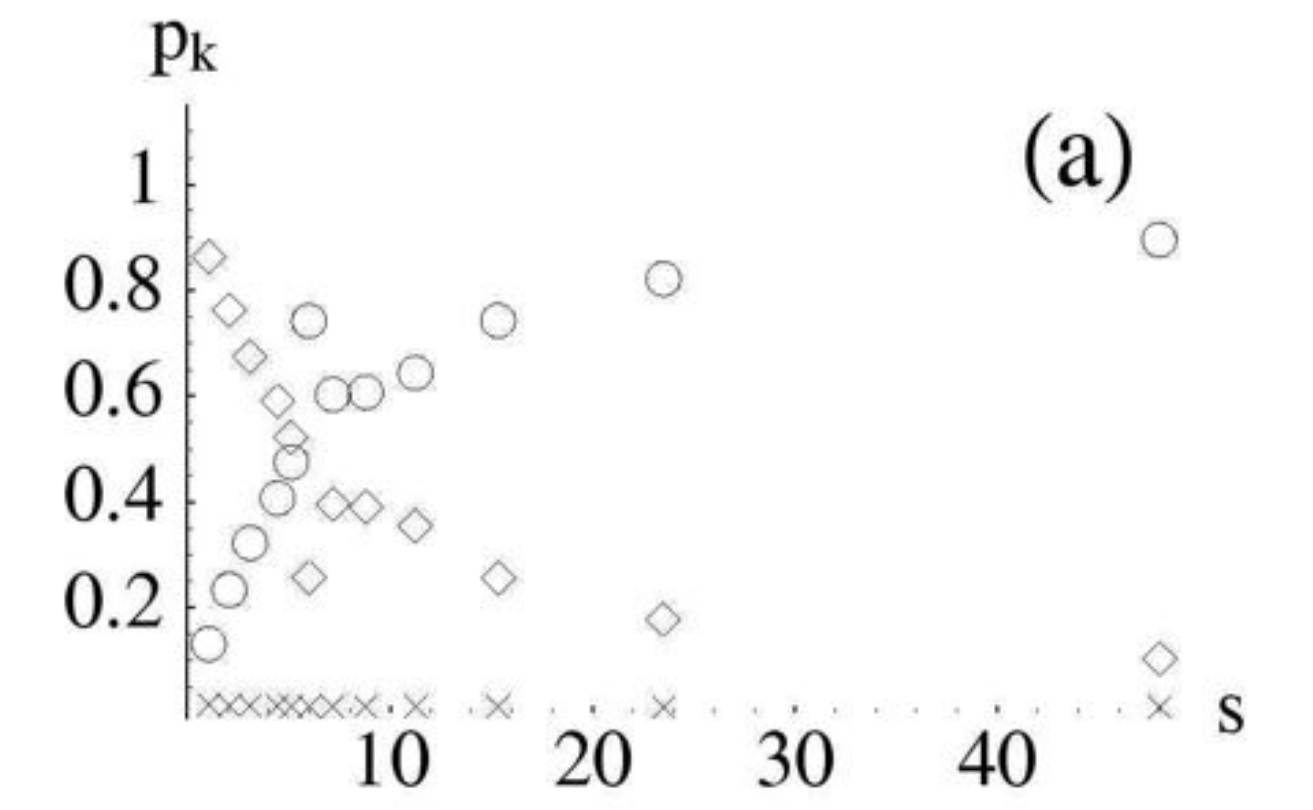}
\includegraphics[width=6. cm]{pRC.pdf}\\
\includegraphics[width=9.5 cm]{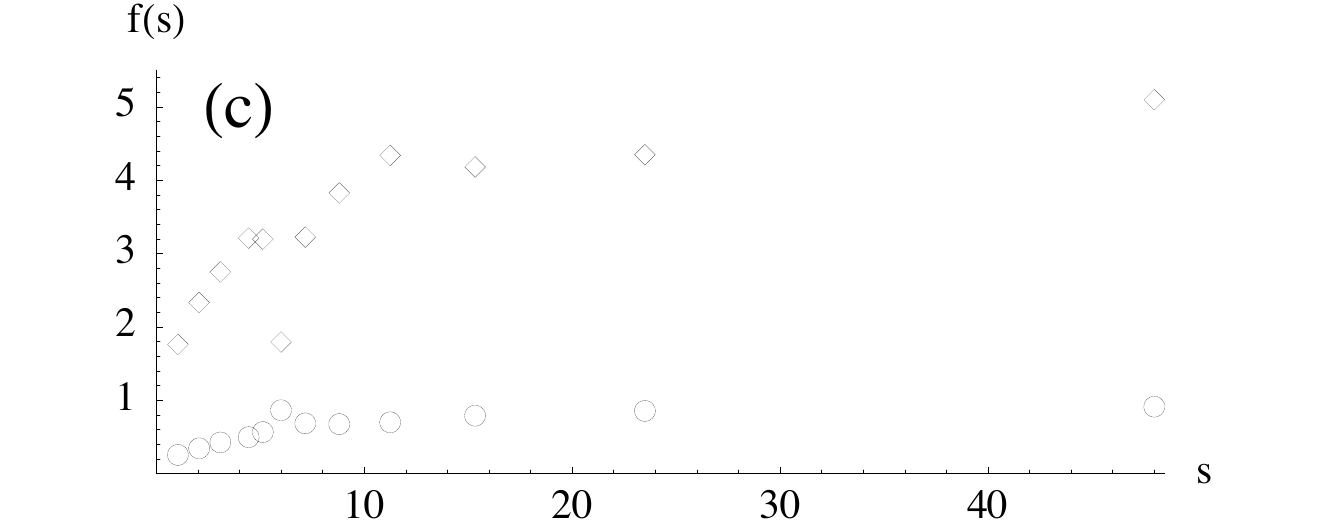}
\caption{In (a) stationary state populations for LH2s (circles), LH1s (diamonds)  and RCs (crosses), as a function of the stoichiometry of membranes presented in Fig.\ref{archs2}. In (b) a zoom of RC populations is made, and in (c) the enhancement factors $f_1(s)$ (diamonds)  and $f_2(s)$ (circles) are presented.}\label{dyns}
\end{figure}

We can also address with use of small architectures the effect of variation in the relative amount of LH1/LH2 complexes, able to change the population of the available states. Fig.\ref{archs2} shows small networks of LH-RC nodes, where the  relative amount of LH2 and LH1 complexes quantified by stoichiometry $s=N_2/N_1$ is varied, in order to study the exciton  dynamics. In Fig.\ref{dyns}(a) the  population ratio at stationary state of LHs demonstrate that as stoichiometry $s$ becomes greater, the population of LH1s, becomes smaller, since their amount is reduced. It is apparent that RC population is quite small, and although their abundance increases the exciton trend to be found in any RC (Fig.\ref{dyns}(b)),  generally, excitations will be found in harvesting complexes.  The population of LHs should be dependent on the ratio of complexes type. As verified in Fig.\ref{dyns}(b), RCs have almost no population, and for the discussion below, they will not be taken into account. Populations can be written  as:
\begin{eqnarray}
\hat{p}_{\mbox{1}}(t\rightarrow\infty)&=& f_1(s) \frac{N_1}{N_1+N_2}=\frac{f_1(s)}{1+s}\label{f1}\\
\hat{p}_{\mbox{2}}(t\rightarrow\infty)&=& f_2(s) \frac{N_2}{N_1+N_2}=\frac{s f_2(s)}{1+s}\label{f2}
\end{eqnarray}
where the dependence on the amount of complexes is made explicit with the ratio $\frac{N_k}{N_1+N_2}$, and where $f_1(s)$ and $f_2(s)$ are enhancement factors. This factor provides information on how the population on individual complexes changes, beyond the features arising from their relative abundance. With use of eqs.(\ref{f1}-\ref{f2}),  $f_1(s)$ and $f_2(s)$ can be numerically calculated provided that $\hat{p}_k(t\rightarrow\infty)$ can be known from the master equation, while $s$ is a parameter given for each network. The results for enhancement factors  are presented in Fig.\ref{dyns}(c). The enhancement factor $f_2(s)$ for LH2 seems to saturate at values below one, as a consequence of the trend of excitations to remain in LH1s. This means that increasing further the number of LH2s will not enhance further the individual LH2 populations.  On the other hand $f_1(s)$ has a broader range, and increases with $s$. This result reflects the fact that population of individual LH1s will become greater as more LH2 complexes surround a given LH1. An unconventional architecture (third column, second row  in Fig.\ref{archs2}) has an outermost line of LH1 complexes, whose connectedness to LH2s is compromised. In all the results in Fig. \ref{dyns} (sixth point), this architecture does not follow the trends just pointed out, as LH1 and RC population,  and enhancement factors, are clearly reduced. The population of LH1 complexes depends on their  neighborhood and connectedness. Whenever connectedness of LH1 complexes is lowered, their population will also be reduced.  Hence, deviations from populations  trend with variation of stoichiometry, are a consequence of different degrees of connectedness of LH1s.

Up to this point, the master equation approach has helped us understand generally the effect of stoichiometry and architecture in small networks. Two conclusions can be made:
\begin{enumerate}
\item{Connectedness of LH2 complexes to LH1s, facilitates transfer to RCs}
\item{The relative amount of LH2/LH1 complexes, namely, stoichiometry $s=N_2/N_1$, when augmented, induces smaller population on LH1-RC complexes. On the other hand, smaller $s$ tends to increase the connectedness of LH1s to LH2s and hence, the population of individual LH1 complexes.}
\end{enumerate}

\subsubsection*{2.2.3 Special pair ionization}
\begin{figure}
\centering
\includegraphics[width=0.45\columnwidth]{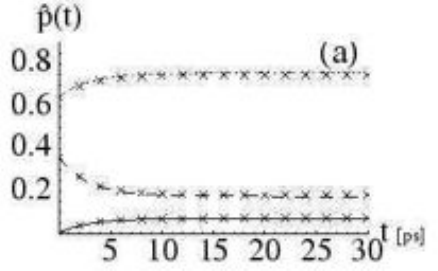}
\includegraphics[width=0.45\columnwidth]{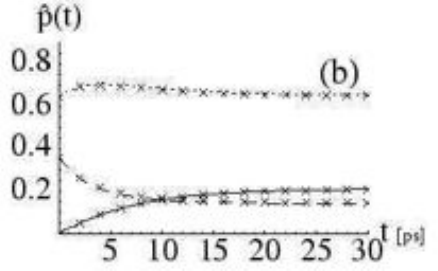}
\caption{Normalized probabilities $\hat{p}_{k}$ for finding the
excitation at an LH2 (dashed), LH1 (dotted)  or at an RC
(continuous), for (a) $t_{+}=3$ps, and (b) $t_{+}\rightarrow
\infty$. Crosses are the results from the Monte Carlo
simulation.}\label{green}
\end{figure}
Another basic process involved in the solar energy conversion is the ionization of the special pair in the RC, and eventual quinol $Q_BH_2$ formation. Remind that once quinol is formed, the special pair is unable to use further incoming excitations before quinol undocks and a new quinone replaces it.  Even though the RC neutrality-diffusion process is propelled by complicated dynamics and involved mechanisms, in an easy approach, let us assume that  the RC dynamics will proceed through a dichotomic process of "open" and "closed" RC states. In the open state, special pair oxidation is possible, while when closed, special pair oxidation to form quinol never happens, hence $t_+\rightarrow\infty$

 The effect of open and closed RC states changes the exciton kinetics. We start with a minimal configuration corresponding to a basic photosynthetic unit: one LH2, one LH1 and its RC.  Figure \ref{green}(a) shows that if the RC is open, excitations will mostly be found in the LH1 complex, followed by
occurrences at the LH2 and lastly at the RC. On the other hand, Figure \ref{green}(b))
shows clearly the different excitation kinetics which arise when the RC is initially
unable to start the electron transfer $P^*\rightarrow P^+$, and then
after $\approx 15$ps the RC population becomes greater with respect to
the LH2's. This confirms that a faithful description of the actual photosynhesis mechanism, even at the level of the minimal unit, must resort into RC cycling, given that its effects are by no means negligible. Moreover, comparison among Figs.\ref{archs0}(d) and \ref{green} also presents a feature that is usually undermined when small architectures are used to straightforward  interpret its results  as truth for  greater, real biological vesicles.  Energy funneling becomes smaller with the number of antenna LH2 complexes, thereby, in architectures with many harvesting antenna complexes, excitation will find it more difficult to arrive to any of the relatively spread RCs. Besides, although LH2$\rightarrow$LH1 transfer rate is five-fold the back-transfer rate, the amount of smaller sized LH2s neighboring a given LH1 will increase the net back-transfer rate due to site availability. Hence, the funneling concept might be  valid for  small networks \citep{review,ritz2001}, however, in natural
scenarios involving entire chromatophores with many complexes,  energy funneling might not be priority due to increased number of available states, provided from all LH2s surrounding a core complex, and globally, from the relative low RC abundance within a real vesicle.

It is important to mention that results for master equation calculations require several minutes in a standard computer to yield the results shown in Fig.\ref{dyns}, and that these networks have an amount of nodes an order of magnitude smaller than the actual chromatophore vesicles. Dynamics concerning the RC cycling have not been described yet, fact that would increase further the dimension of possible membrane's states. To circumvent this problem, further analysis will proceed from {\it stochastic simulations}, and observables will be obtained from ensemble averages.

\subsection*{2.3 Full vesicles}
A real vesicle involves several hundreds of harvesting complexes. Given the large state-space needed to describe such amount of complexes and our interest  to inquire on a variety of incoming light statistics in the sections ahead, our subsequent model analysis will be based on a discrete-time random walk
for  excitation hopping between neighboring complexes.
\subsubsection{Simulation algorithm.}\label{simulations}
In particular, we use a Monte Carlo method to simulate the events of  excitation  transfer,
the photon absorption, the dissipation,  and the RC
electron transfer. We have checked that our Monte Carlo simulations accurately reproduce the results of the population-based calculations described above,
as can be seen from Figs.\ref{green}(a) and (b).
The Monte Carlo simulations proceed as follows. In general, any distribution of light might be used with the restriction of having a mean inter-photon time of $\gamma_A^{-1}$ from eq.(\ref{gammaa}). Accordingly, a first photon is captured by the membrane and the time for the next absorption is set by inverting the cumulative distribution function from a [0,1] uniformly distributed (Unit Uniformly Distributed, UUD) random  number. This inversing procedure is used for any transfer, dissipation or quinol removal event as well. The chosen absorbing complex is randomly selected first among LH1 or LH2 by a second UUD number compared to the probability of absorption in such complex kind, say $N_{1(2)}\gamma_{1(2)}/\gamma_A$ for LH1 (LH2), and a third UUD random number to specifically select any of the given complexes,  with probability $1/N_{1(2)}$. Once the excitation is within a given complex, the conditional master equation given that full knowledge of the excitation residing in site $i$, only involves transfers outside such site, say $\partial_t\rho_i=-(\sum_j 1/t_{i,j}+\gamma_D)\rho_i$, whose solution is straightforward to provide the survival probability and its inverse,  of use to choose the time $t^*$ for the next event according to eq.(\ref{meq}) from a UUD number $r$: $-\log r/(\sum_j1/t_{i,j} +\gamma_D)=t^*$.  Once $t^*$ is found, a particular event is chosen: transfer to a given neighboring complex $j$ with probability $(1/t_{i,j})/(\sum_j 1/t_{i,j}+\gamma_D)$ or dissipation with probability $\gamma_D/(\sum_j 1/t_{i,j}+\gamma_D)$, which are assigned a proportional segment within [0,1] and compared with another UUD number to pinpoint the particular event. If the chosen event is a transfer step, then the excitation jumps to the chosen complex and the transfer-dissipation algorithm starts again. If dissipation occurs, the absorption algorithm is called to initiate a new excitation history. In a RC, the channel of quinol ionization is present with a rate $1/t_+$ in an event that if chosen, produces the same effect as dissipation. Nonetheless, the number of excitations that become SP ionizations are counted on each RC, such that when two excitations ionize a given RC and produce quinol, it becomes closed by temporally setting $1/t_+=0$ at such RC.   Quinol unbinding will set ``open" the RC, not before the RC-cycling time with mean $\tau$, has elapsed, chosen according to a poissonian distribution.
The algorithm can be summarized as follows:
\begin{enumerate}
  \item {Create the network: Obtain coordinates and type of LHs, and label complexes, for instance, by solely numerating them along its type, say complex 132 is of type 2 (we use 1 for LH1, 2 for LH2 and 3 for RC). Choose the $j$ neighbors of complex $i$ according to a maximum center to center distance less than $r_1+r_2+\delta$, $r_2+r_2+\delta$ and $r_1+r_1+\delta$ for respective complexes. We use $\delta=20\AA$, chosen such that only nearest neighbors are accounted and further increase of $\delta$ makes no difference on the amount of nearest neighbor connections, although further increase may include non-physical next to near-neighbors. In practice, the network creation was done by three arrays, one, say $neigh(i,j)$ with size $M\times S$, with $M$ complexes as described above, and $S$ as the maximum number  of neighboring complexes among all the sites, hence requiring several attempts to be determined. Minimally $j\le1$ for an LH2, concerning the dissipation channel, $j\le2$ for LH1 including both dissipation and transfer to its RC, and $j\le3$ for a RC accounting on dissipation, RC ionization an transfer to its surrounding LH1.  The other arrays are built, say $size(i)$, with $M$ positions, that keep on each the number of neighbors of the respective $i$ labelled complex, and $rates (i,j)$ where at each position the inter-complex rate $i\rightarrow j$ is saved.  For instance, $rates(i,1)$ of any RC will be the ionization rate $1/t_+$.}
 \item {Send  photons to the network: On a time $t^*=-\log(r)/\gamma_A$ according to eq.(\ref{gammaa}), with $r$ being an UUD number. Choose an LH2 or an LH1, according to the probability of absorption from the cross section of complex type  $N_{1(2)}\gamma_{1(2)}/\gamma_A$.  Add one excitation to the network, say $n=n+1$, and assign the initial position $pos(n)=i$ of the excitation according to another UUD that selects an specific labelled $i$ complex. Remind that $n$ is bounded by the maximum amount of excitations allowed to be at the same time within the membrane, usually being one.}\label{absorp}
\item {If the $i$th complex is excited, the construction of the above mentioned arrays make the cycle of excitation dynamics straightforward since the network is created only once, and dynamics only require to save the complex $i$ where the excitation is, and then go through cycles of size $size(n,i)$ to acknowledge the  stochastically generated next time for a given  event. Excitation  can be transferred to the available neighbors, become dissipated or a RC ionization event. Order all times for next events in order to know which will be the next in the array, say $listimes(p)$ with $p\le n$, where $t_{\mbox{min}}=listemp(1)$. In parallel, update an organized array that saves the next process with the number of the neighbor to which hopping occurs, or say a negative number for RC ioinization and another negative number for dissipation.}
\item{Jump to next event: By cycling over the $n$ present excitations, increase time up to the next event $t_{\mbox{min}}$. If RC cycling is accounted, check which time among $t_{min}$ and the next opening RC time $t_{RC}$ (its algorithm is to be discussed in the following) is the closest, and jump to it. }\label{next}
\item{Change state of excitations or that of RCs: Update the current site of the excitation $n$, or whether it becomes a dissipation or a RC ionization. If the latter process occurs, keep in an array, say $rcstate(k)$ whose size equals the total amount of RCs, the number of excitations that have become ionizations from the last time the $k$th RC was opened.If $rcstate(k)=2$ then the $k$th RC is closed by redefinition of $rate (i,1)=0$ and a poissonian stochastically generated opening time with mean $\tau$ is generated. This time interval is kept in an array $rctimes(k)$. Now, introduce this time interval into an ascending ordered list among all closed RCs opening times such that the minimum $t_{RC}$ is obtained. If $t_{RC}< t_{\mbox{min}}$ then jump to that time and open the $k$th RC by letting $rcstate(k)=0$ and $rates(i,1)=1/t_+$.}
\item{Look which is minimum among $t^*$, $t_{RC}$ and $t_{\mbox{min}}$ and jump to steps \ref{absorp} or \ref{next} according to whether $t^*<(t_{RC},t_{\mbox{min}})$ or $(t_{RC},t_{\mbox{min}})<t^*$, respectively.}
\item{If the maximum amount of excitations chosen from the initialization, have been sent to the membrane, finish all processes and write external files.}
\end{enumerate}

The language used to program this algorithm was FORTRAN77,  to point out that these calculations do not require any high-level language.

\subsubsection{ Excitation dynamics trends in many node-complexes networks.}
\begin{figure}
\centering

\includegraphics[width=0.32\columnwidth]{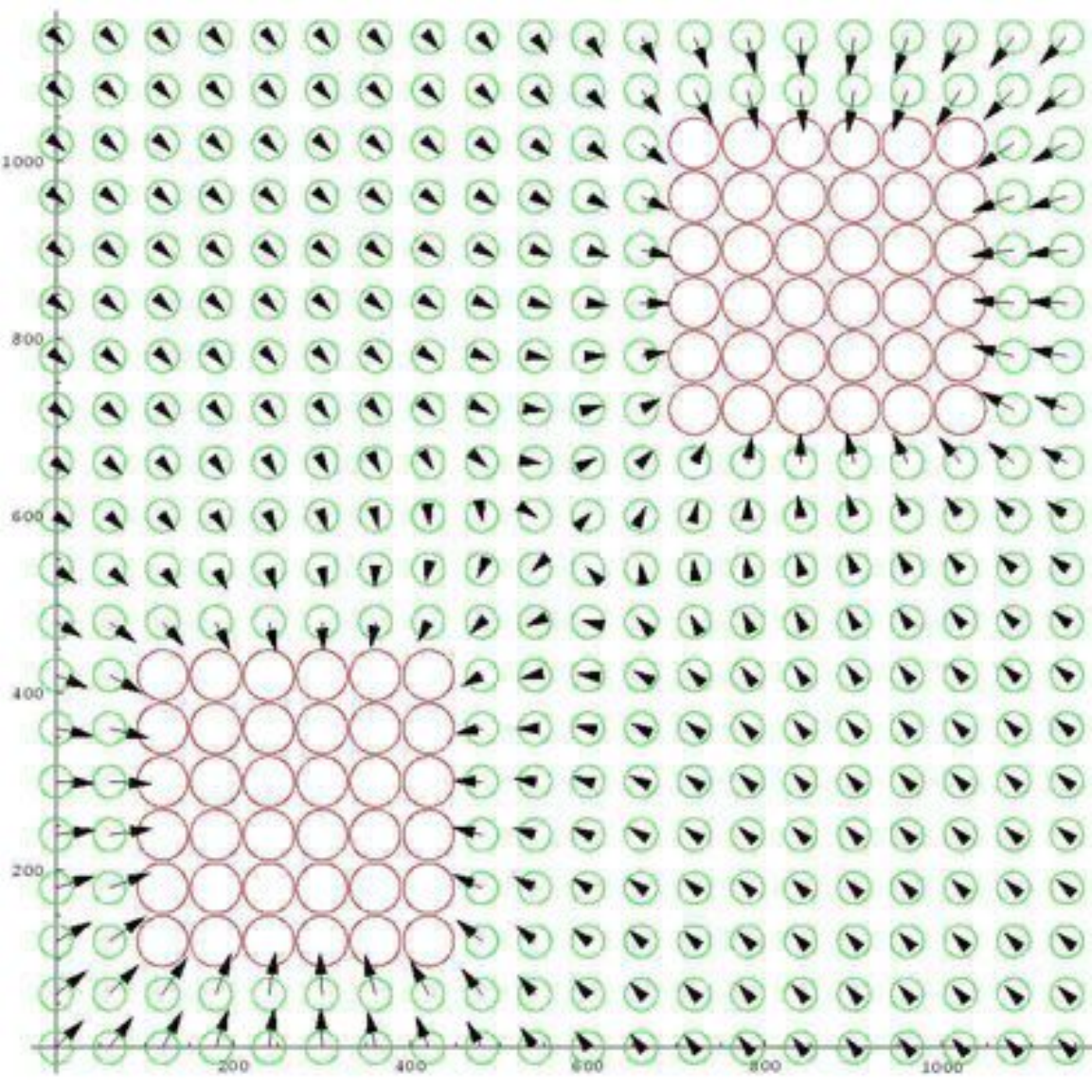}
\includegraphics[width=0.32\columnwidth]{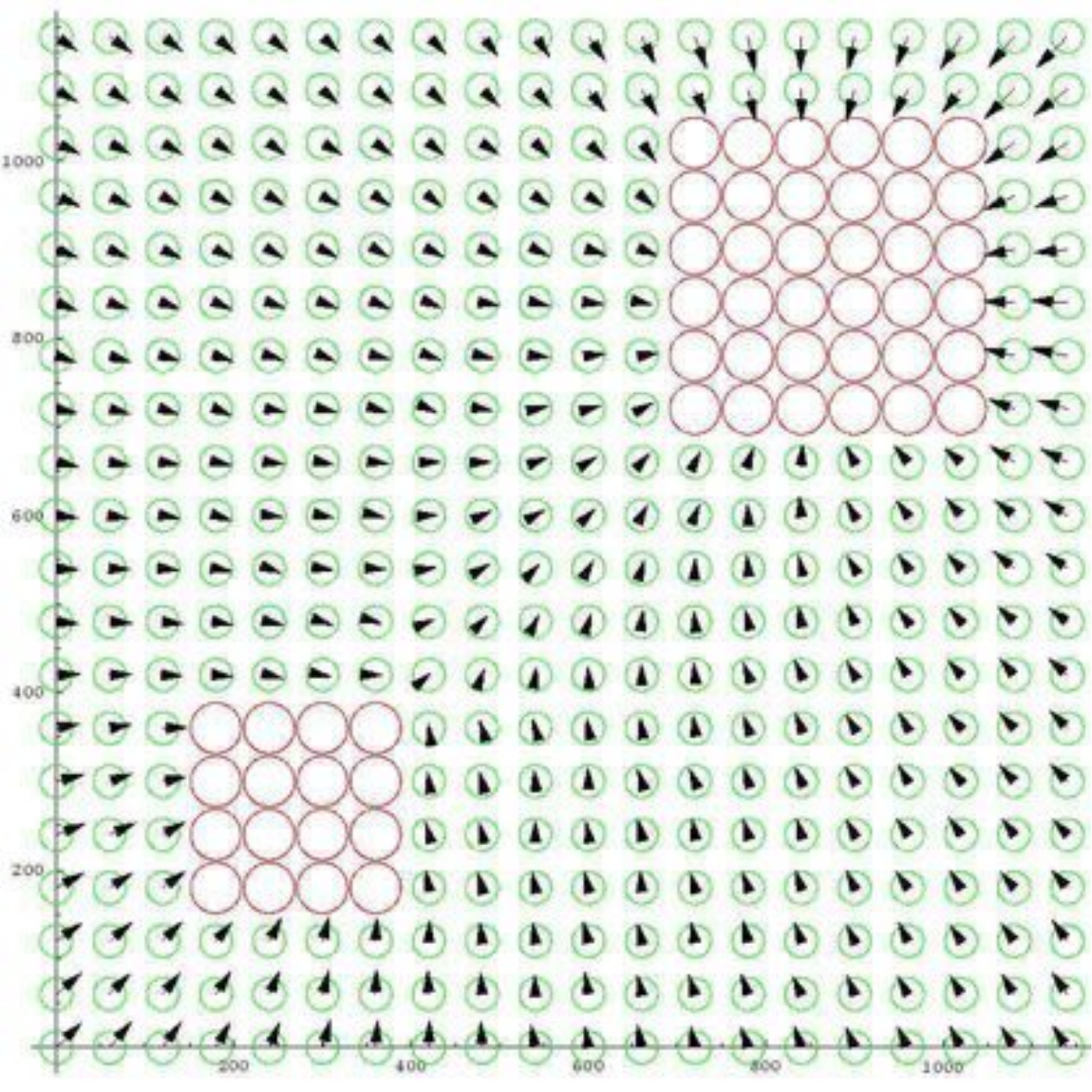}
\includegraphics[width=0.32\columnwidth]{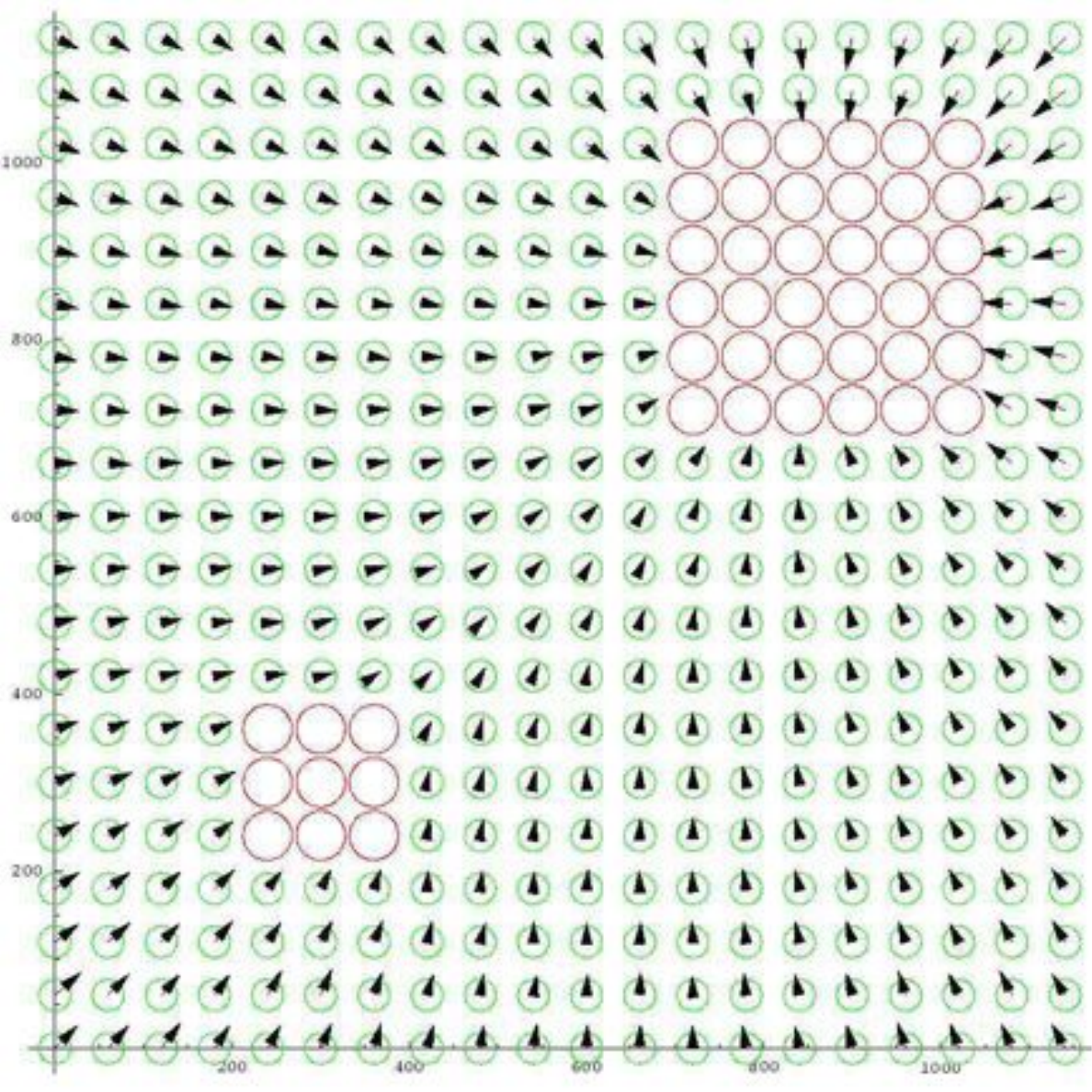}\\
\includegraphics[width=0.32\columnwidth]{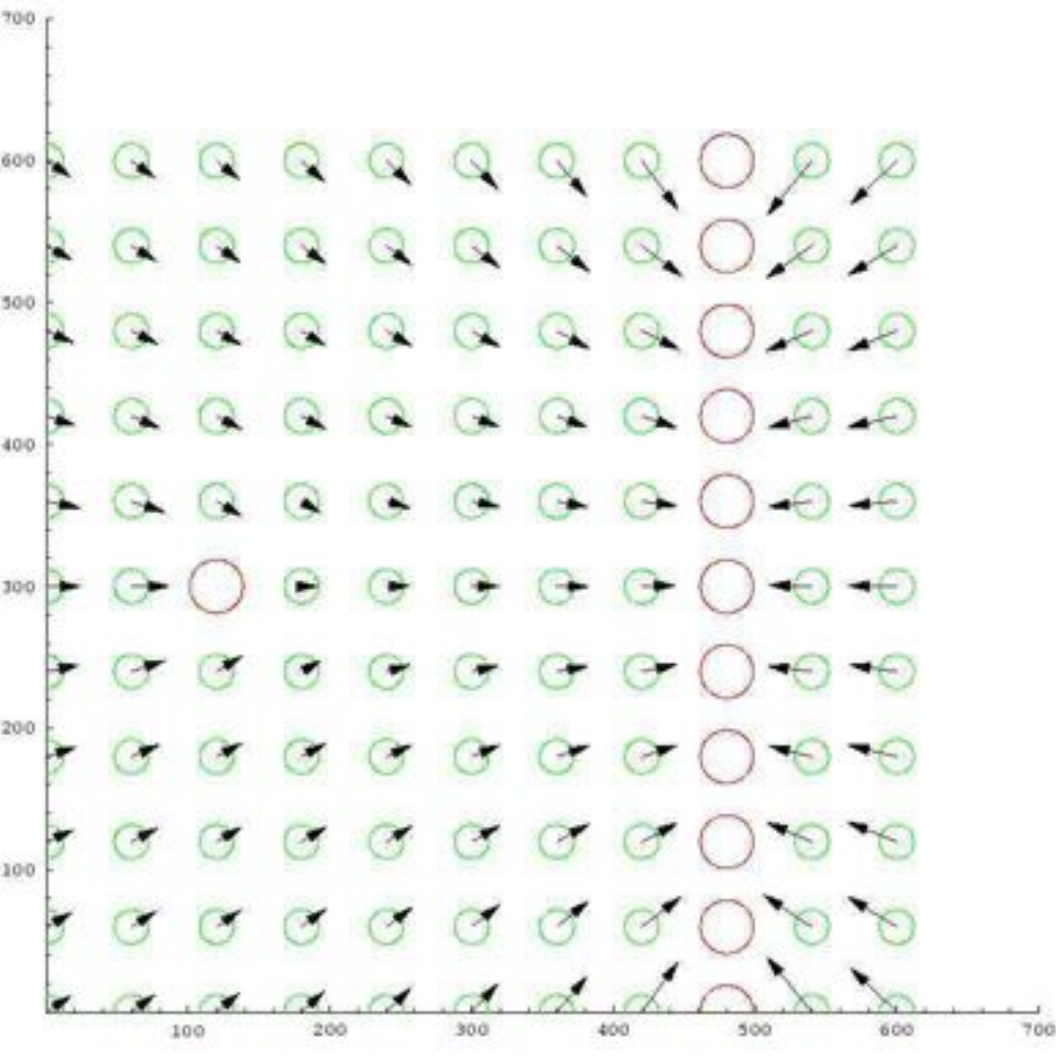}
\includegraphics[width=0.32\columnwidth]{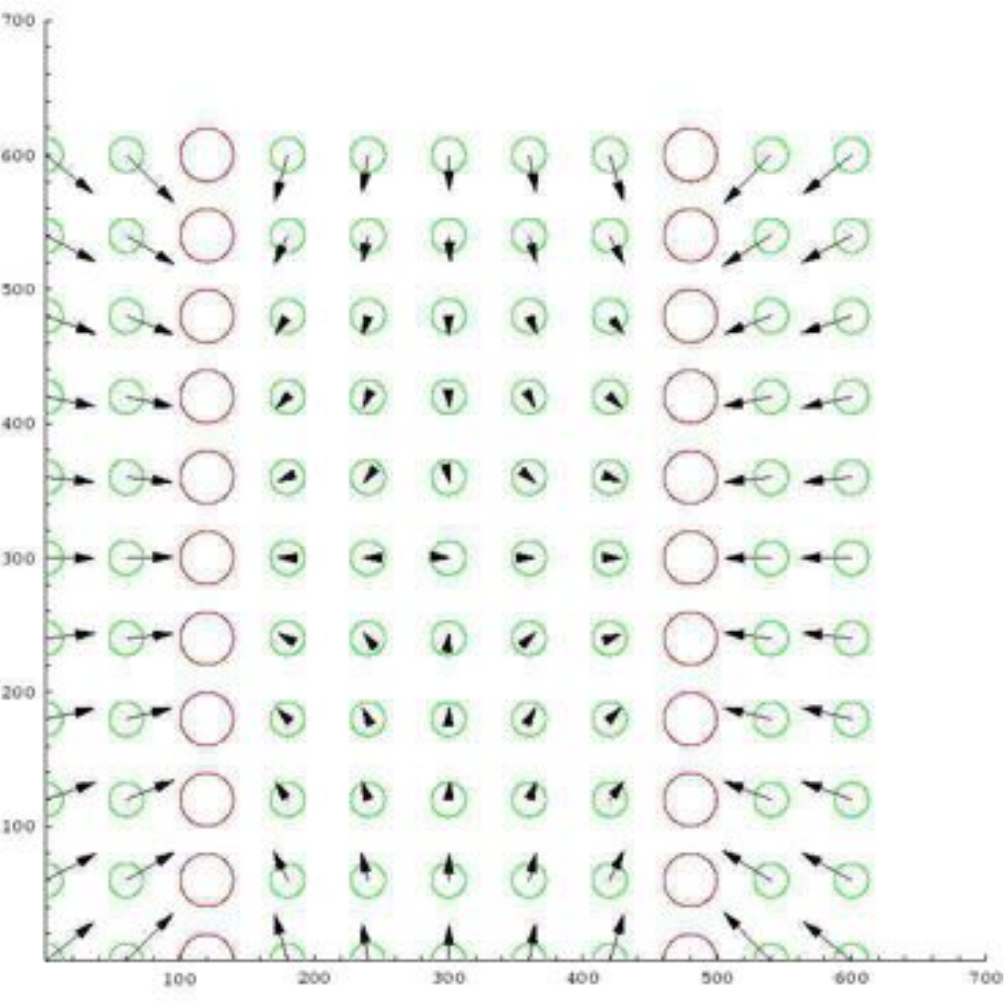}
\includegraphics[width=0.32\columnwidth]{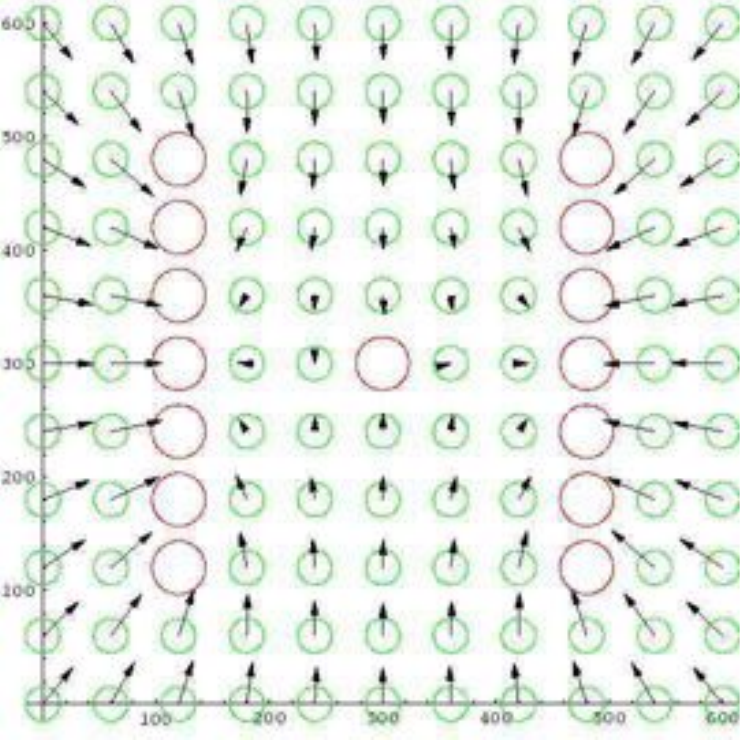}\\
\includegraphics[width=0.32\columnwidth]{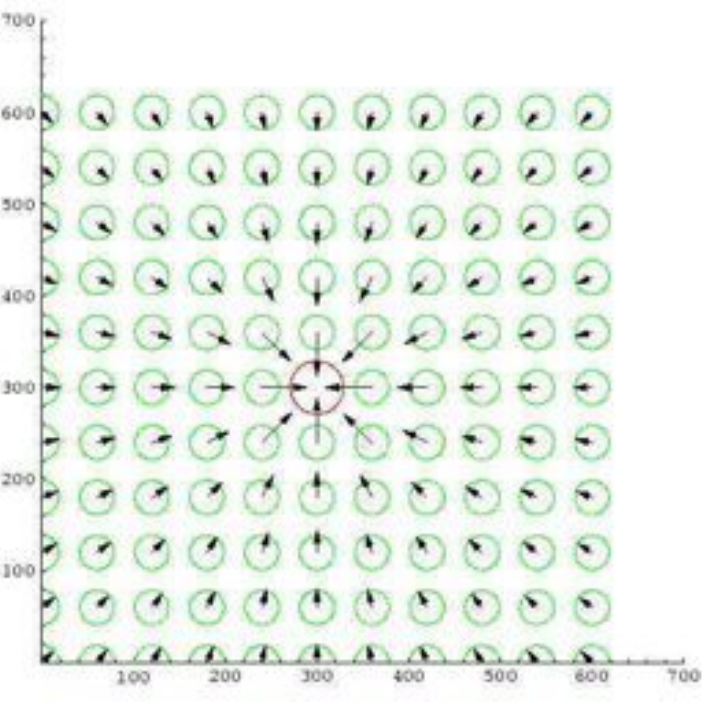}
\includegraphics[width=0.32\columnwidth]{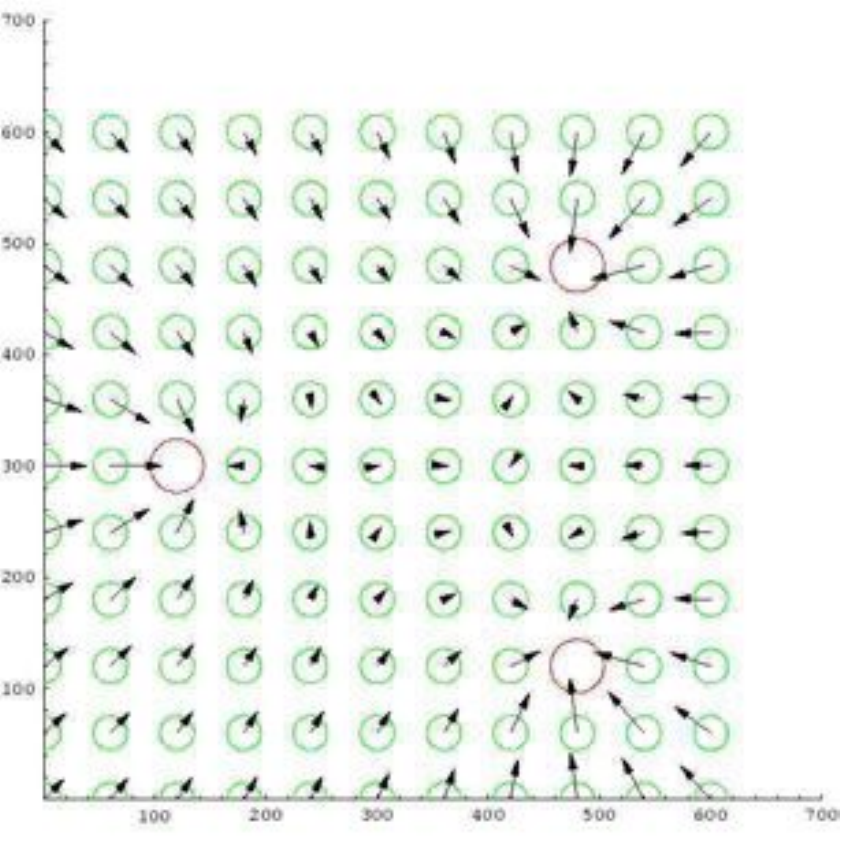}
\includegraphics[width=0.32\columnwidth]{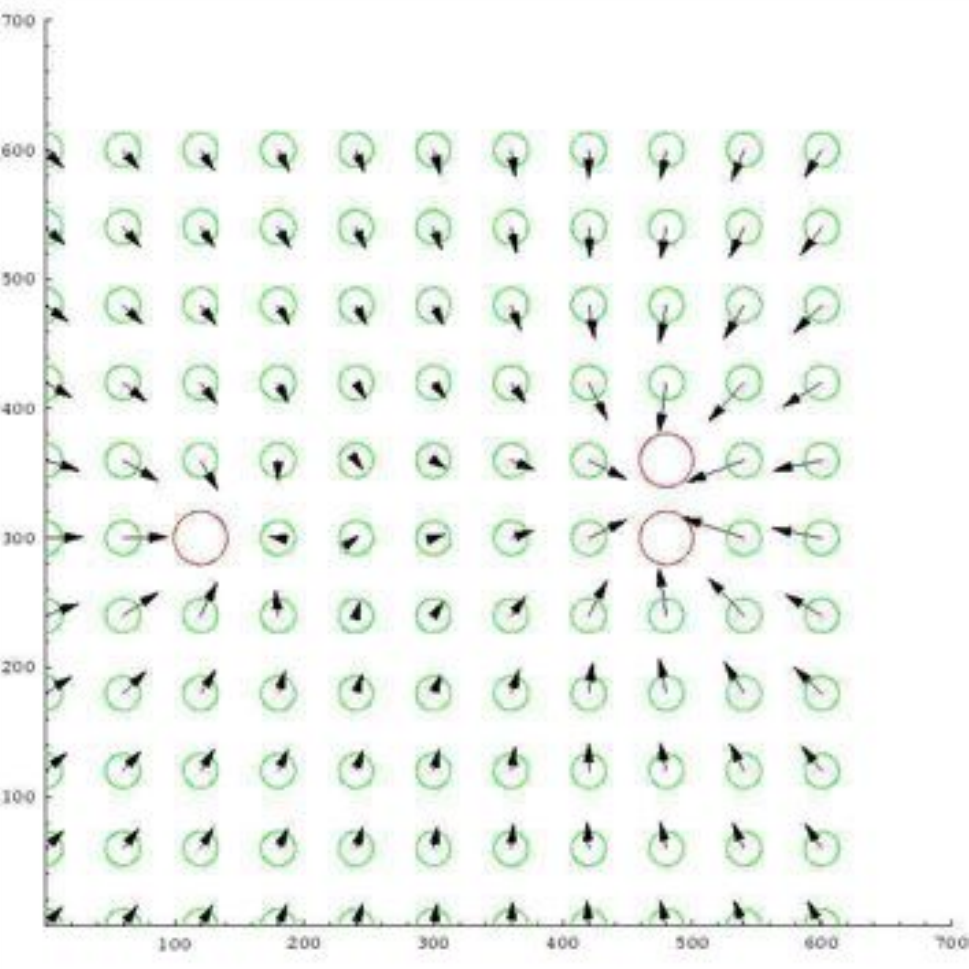}\\
\caption{Vector field ${\bf v}$ for different configurations, where the effect of  core centers clustering is investigated (100000 runs starting in each site).}\label{toyfull}
\end{figure}

In order to understand at a qualitative degree the excitation dynamics trends involved in  full network chromatophores, a few toy architectures have been studied, shown in
Fig.\ref{toyfull}. In this preliminary study it is of interest to understand the excitation kinetics in complete chromatophores. In particular, it is useful to understand if any important feature arises according to nature's found tendency of forming clusters of the same complex type. In AFM images \citep{scheuring2005} it has been found that there is an apparent trend to form clusters of LH1 complexes with simultaneous formation of LH2 para-chrystalline domains. The reason that has been argued for this trend \citep{scheuring2005} involves the RC cycling dynamics and can be explained as follows.

The charge carrier quinone-quinol require diffusing through the intracytoplasmic membrane within the void spaces left among harvesting complexes, in order to reach the bc1 complex and complete the electric charge cycle with cytochorme that however, diffuses through the periplasm as schematically shown in Fig.\ref{fig2}. The closeness among LH2 complexes in these para-chrystalline domains restricts the void spaces required for diffusion of quinone-quinol to the LH1 domains, where charge separation is taking place. Then, such aggregation indeed  helps to improve the time it takes to quinone perform the whole RC-bc1-RC cycle, by restricting its presence to RC domains. However, an advantage concerning excitation dynamics has been heuristically proposed, where the path an excitation has to travel to reach an open RC once it encounters a closed RC is reduced due to LH1 clustering. In this section we investigate this latter possibility from a more quantitative point of view and address the former in the next section.

In order to understand exciton kinetics, for instance, let us fix our attention on the probability of a RC ionization when excitations start in a given LH2 on the configurations shown in Fig.\ref{toyfull}. To that end, let us introduce first the number $n_{RC{ij}}$ of excitations that are absorbed at site $i$ and become ionization of the special pair at a given RC, labeled $j$. Also let $\hat{r}_{ij}$ be the unitary  vector pointing in the direction from LH2 site $i$ to RC site $j$. As an analogous to a force field, let  vector field ${\bf v}$ correspond to a weighted sum of the directions on which RC ionizations occur, starting for  a given LH2 complex
\begin{equation}
{\bf v}_i=\frac{\sum_j n_{RC_{ij}}\hat{r}_{ij}} {\sqrt{\sum_j n_{RC_{ij}}^2}}.
\end{equation}
of help as a purely numerical calculation of graphical interpretation. The normalization gives a basis to compare only the directions for RC ionizations independent whether excitations are dissipated or not. Note that $\hat{r}_{ij}$ has both positive and negative signs, hence the vector ${\bf v}_i$ will have greater magnitude as RC ionizations occur more frequently at a given direction, or less magnitude as ionizations occur in more spread out directions.

Several features can be understood from this vector field. Figure \ref{toyfull}  shows the vector  fields  attracted to the LH1 most abundant  regions. The arrows illustrate the favored statistical direction in which excitations will become RC ionization, and they do not imply that an excitation will deterministically move in that direction  if it were absorbed in such place. Firstly, in all cases,  a tendency to point on clustered LH1s stands from  distant LH2 complexes, a fact that emphasizes the very rapid transfer rates that lead to excitation's rapid global membrane sampling. Secondly, it is evident that better excitation attractors are made with increasing the core centers cluster size, easily observed in the top panel. Hence,  according to this result one may naively state that a funneling effect is apparent and strongly dependent on cluster size. However, on the middle panel, it can be seen that this funneling effect is vanishingly small at sites where  some degree of symmetry is present concerning the distribution of neighboring RCs, even though if these latter display some clustering degree. Third, a trend is displayed where the spreading of the flux becomes smaller with a lesser amount of RCs. However it is important to be aware that in more detail and in a general fashion, one also sees that even the  LH2s closest  to a given RC, "feel" the presence of further RCs. This implies that a significant amount of excitations that reach LH2 complexes neighboring a given LH1, will better prefer hopping to neighboring LH2 complexes to eventually reach a distant RC: no "funneling", as usually understood.

\begin{figure}
\centering
\includegraphics[width=0.47\columnwidth]{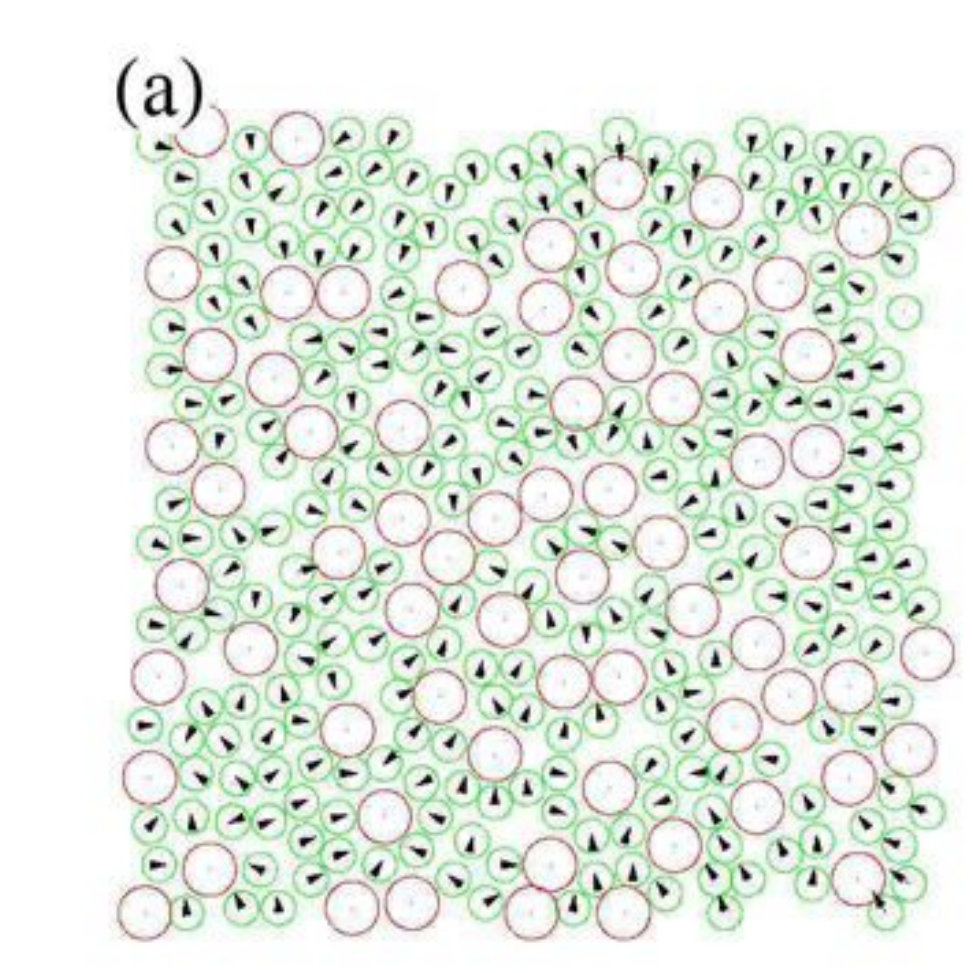}
\includegraphics[width=0.47\columnwidth]{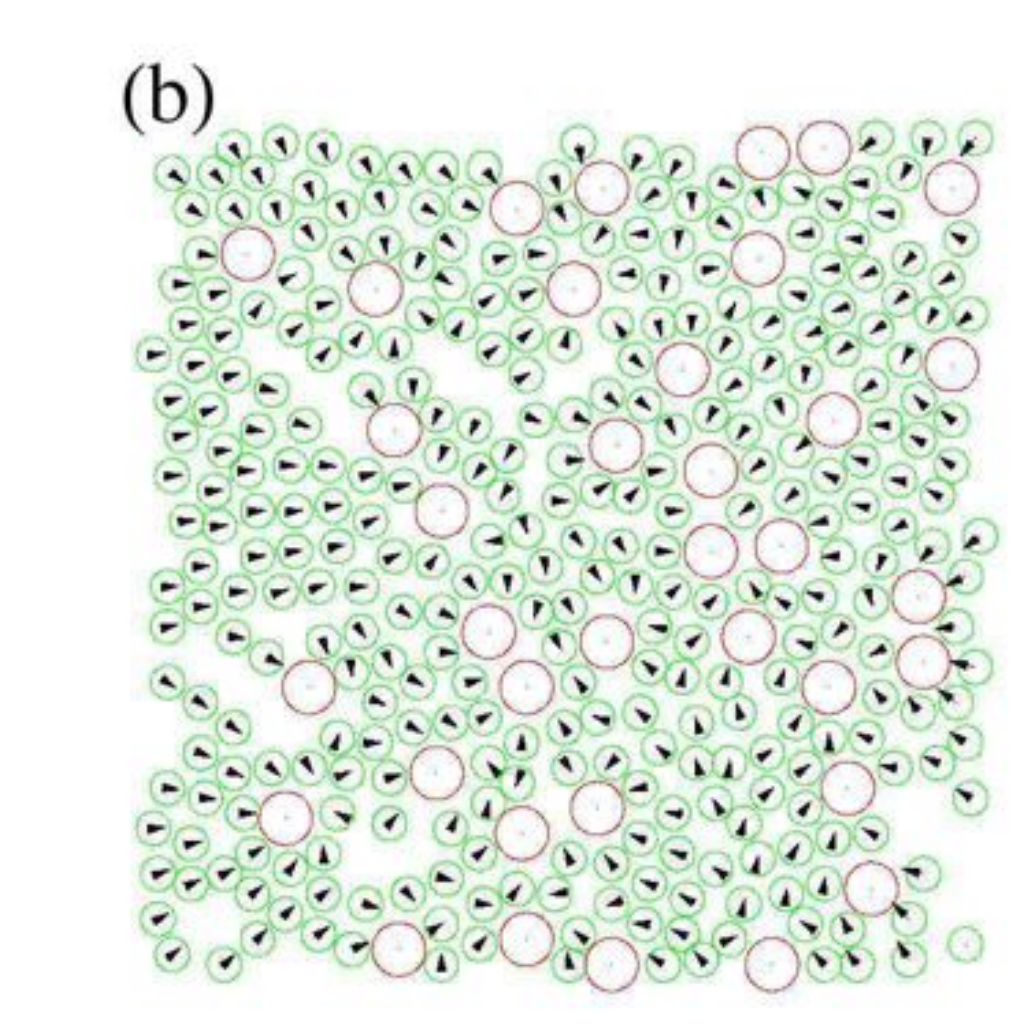}
\caption{Vector field ${\bf v}$ for HLI (a) and LLI (b) membranes (100000 runs each site).}\label{realarch}
\end{figure}

In a more realistic situation, the clustering of LH1 complexes is just a trend, and in average any cluster is formed by a few 3-4 core complexes in HLI situations and somehow less for LLI membranes as shown in  Fig.\ref{realarch}, where empirical architectures are presented in complete accordance with experimental data, taken from Ref.\citep{scheuring2004}.  Along these figures, the vector field calculated for the HLI and LLI  empirical membranes is made, to highlight no clear trend of excitations to be transferred  immediately to neighboring LH1 complexes, and disregard a funneling LH2$\rightarrow$LH1 as the unique direction of excitation flux when many complexes are accounted.

In detail, from the reasons just provided, in  Fig.\ref{realarch}(a) it is apparent that in HLI membranes, individual LH1 complexes are not efficient attractors of excitations.  Here, not even the LH2s closest to a core center complex are notably attracted to such cores. On the other hand, a more uniform flux seems to be shared on clusters of LH2s in the  LLI membrane (Fig.\ref{realarch}(b)), that points to locations where more core centers are displayed.  Hence, it is more evident the directionality of the flux to the most  LH1 abundant locations that gives a statistical preference to ionize the closest RCs due their low global abundance in LLI membranes.  The greater amount of LH1 complexes in the HLI membrane induces no significant preference for excitations to hop to the closest lying RCs and provides evidence that the excitation process proceeds without a clear funneling effect. If a single excitation can become an ionization in a single RC, the flux is clear, but if there are many and spread, the flux becomes random, as would be expected from any random walk.  In summary, these results allow a mental picture of the excitation transfer process: Excitations become absorbed and start wandering along the membrane in a random walk regardless of a funneling effect, up to the moment where they reach a RC and become ionized, in a process where statistical preference is present in LLI membranes due to the reduced amount of RC, and presenting more clearly the random wandering on HLI due to higher spread out possibilities of ionization. This fact can also be quantitatively investigated.

 Suppose that a given LH1 is surrounded  by a few LH2 complexes. If an excitation is absorbed into one of these neighboring LH2s, it will have a survival probability decaying with rate $1/t_{21}+n/t_{22}$ where $n$ wold be 3-4 corresponding to the number of LH2s neighboring such absorbing LH2. Given the presented transfer times (remind $t_{21}=3.3$ps and $t_{22}=10$ps) both terms in the survival decay rate become almost equal for $n=3$, and hence no preference will occur on the excitation to be transferred to the closest LH1. If LH2$\rightarrow$LH1 transfer would actually happen, the same exercise can be done for the survival probability within the LH1 complex, decaying with a rate $m/t_{12}+1/t_{1,RC}$    (remind $t_{12}=15.5$ps and $t_{1,RC}\approx30$ps) which assuming a single surrounding LH2 complex  ($m=1$)  would give a preference for excitations to return to antenna complexes. Hence, if the availability of neighbors is accounted, the funneling  LH2$\rightarrow$LH1 is lost, which along the enhanced back-transfer rate LH1$\rightarrow$LH2 as compared to LH1$\rightarrow$RC, provides the basis for understanding complete chromatophores as networks where actual performance depends upon the RC availability, since the event of ionization only depends on the probability to get to RC sites having no preference to be visited within the network. Excitations sample great portions of the whole membrane in its hopping, hence become able to reach RCs far away from the absorption sites before dissipation overcomes.

Dissipation itself also provides  interesting features.  The dissipation $d_i$ measures the probability for  excitations to dissipate at site $i$ from the the total amount of excitations being absorbed in the membrane $n_A$:
\begin{equation}
d_i=\frac{n_{D_i}}{n_A}
\end{equation}
that has an straightforward relation with the global efficiency $\eta$ of the membrane, that accounts on the probability of any excitation to be used as an SP ionization. Given that excitations can only become RC ionizations  or be dissipated, the sum over all complexes of dissipation probability will give the probability of any excitation to be dissipated, hence
\begin{equation}
\eta=1-\sum_i d_i
\end{equation}

\begin{figure}
\includegraphics[width=0.45\columnwidth]{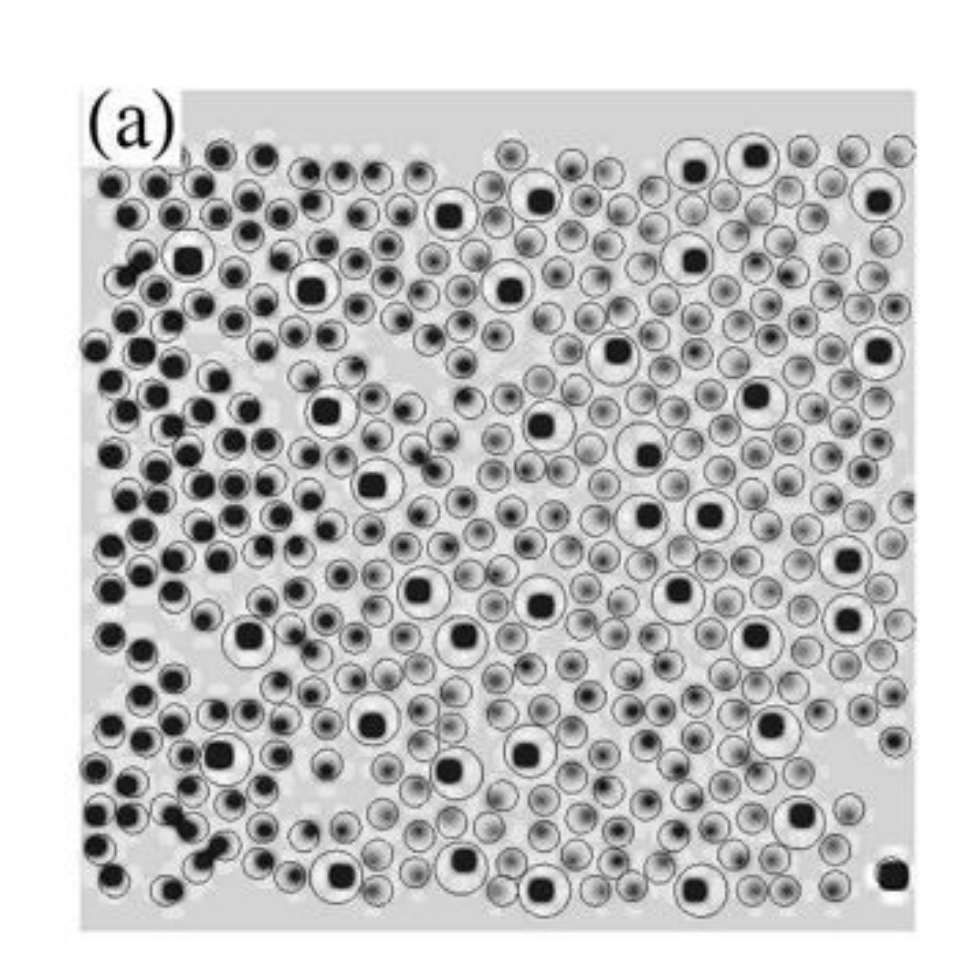}
\includegraphics[width=0.45\columnwidth]{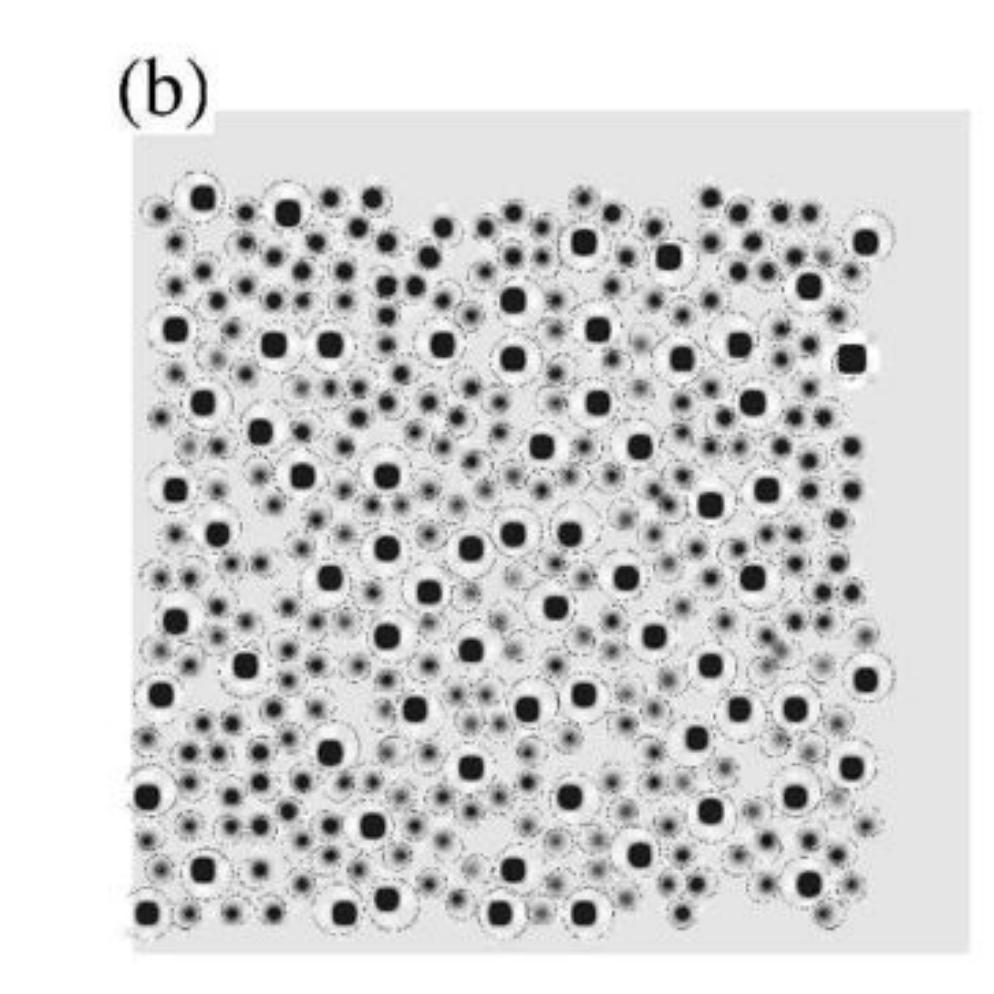}
\caption{Contour plots for dissipation in  LLI (a) and HLI (b)
membranes. Greater contrast means higher values for dissipation.
The simulation is shown after $10^6$ excitations were absorbed by
the membrane with rate $\gamma_A$.}\label{dissmembranes}
\end{figure}

Figure \ref{dissmembranes} shows the dissipation on the membranes shinned with the respective illumination rate of their growth (LLI 10 W/m$^2$ and HLI 100 W/m$^2$). Figure \ref{dissmembranes}(a) shows that the LLI membrane  has highly dissipative clusters of LH2s, in contrast to the uniform dissipation
in the HLI  membrane (see Fig.\ref{dissmembranes}(b)).

This result addresses distinct features probably connected with the own requirements for bacteria when different light intensity is used during growth.  Under LLI, bacterium might require use of all the available solar energy to promote its metabolism. In the numerical simulations (as expected from nature), the dissipation rate is set equal on any site, and therefore dissipation is only dependent on the probability for an excitation to be in a given complex. If more dissipation is found in some regions, it can there be supposed that excitations remain more time at such domains. Hence the tendency for excitations to dissipate implies they reside longer in  LH2 complex domains and justify the view of LH2 clusters as excitation reservoirs. Although RC cycling has not been accounted yet, a given history might include an event where the excitation reaches closed clustered RCs and then jumps back to these LH2 domains  waiting before the RCs become available again. On the other hand, HLI membranes display an evenly distributed dissipation. For instance, at HLI not all excitations might be required and dissipation can be used as a photo-protective mechanism. In this case, if dissipation were highly concentrated, vibrational recombination would overheat such highly dissipating domains.

Beyond these local details, an analysis regarding average values of dissipation  $D_k$ and residence  $p_{R_k}$ probabilities  on a complex type $k$ ($k=1,2$ are respectively LH1, LH2 complexes) better supports the view of a completely random excitation hopping process. Table \ref{table1} shows the numerical results of $p_{R_k}$, that concerns the probability to find an excitation in a given complex type and is calculated from the sum of the residence  times  of excitations at a given complex type. It is straightforward to see that $p_{R_k}$ is closely related to the probability of dissipation at such complex type, therefore, dissipation can correctly measure where excitations are to be found in general.  The randomness of excitation dynamics  is illustrated by realizing that dissipation in a given complex type depends primarily
on its relative abundance, since  $\frac{D_{k}}{D_j}\approx\frac{N_k}{N_j}$, and justifies that apart from LH2 clusters local variations, in the mean, all sites behave equivalently, and no dynamical feature arises to set a difference among LH1s and LH2s able to  argue on the stoichiometry adaptation experimentally found. Lastly, as expected from the dynamical equivalence of sites, along the result obtained with toy architectures of varying stoichiometry where RC population heavily rose due their abundance but relatively invariant to arrangement, the probability to reach a RC solely depends on its abundance, and therefore the global efficiency $\eta$ is greater in HLI membranes.

\begin{table}[hh]
\centering
\begin{tabular}{|r|r|r|r|r|r|r|r|}
\hline
Membrane & $p_{R_2}$ & $p_{R_1}$& $D_{2}$ & $D_{1}$  & $\frac{D_{2}}{D_{1}}$ & $s=\frac{N_2}{N_1}$&$\eta=1-\frac{n_{D}}{n_A}$\\
\hline
LLI & 0.72& 0.25 &0.74 & 0.26 & 9.13 & 9.13 &0.86\\
HLI &0.50 & 0.46 & 0.52 & 0.48 & 3.88 & 3.92 &0.91\\
\hline
\end{tabular}
\caption{Dissipation $D_k$, residence probability $p_{R_k}$,
on $k=\{1,2\}$ corresponding to $N_1$ LH1 and $N_2$ LH2 complexes
respectively. Stoichiometry $s$ and efficiency $\eta$ are also
shown.}\label{table1}
\end{table}
Hence, for the present discussion, the most important finding from these numerical simulations is that the adaptation of purple bacteria
does {\em not} lie in the single excitation kinetics. In particular, LLI
membranes are seen to reduce their efficiency globally at the point where photons are becoming
scarcer -- hence the answer to adaptation must lie in some more fundamental trade-off (as we will later show explicitly). Due to the dissimilar timescales between millisecond absorption from eq.(\ref{gammaa}) and nanosecond
dissipation,  multiple excitation dynamics are also unlikely to occur within a membrane. However, multiple excitation dynamics cannot be regarded {\it a priori} not to be a reason for purple bacteria adaptation. A numerical study in Ref.\citep{NJPCaycedo-Soler} involving  blockade in which two excitations can not occupy the same site and annihilation, where two excitations annihilate due to vibrational recombination occurring due to significant Frank-Condon overlap on higher exciton states, shows that  these mechanisms decrease the efficiency of membranes equally on LLI and HLI membranes, therefore, keeping the best performance to HLI. 

Nevertheless a more comprehensive study to be presented  in the next section taking into account the coupled exciton-RC cycling is able to justify nature's choice of proceeding with the observed bacterial adaptations.

\section*{3. Complete chromatophores: Exciton and RC cycling coupled dynamics.}\label{vesicles}

\begin{figure}
\centering
\includegraphics[width=0.65\columnwidth]{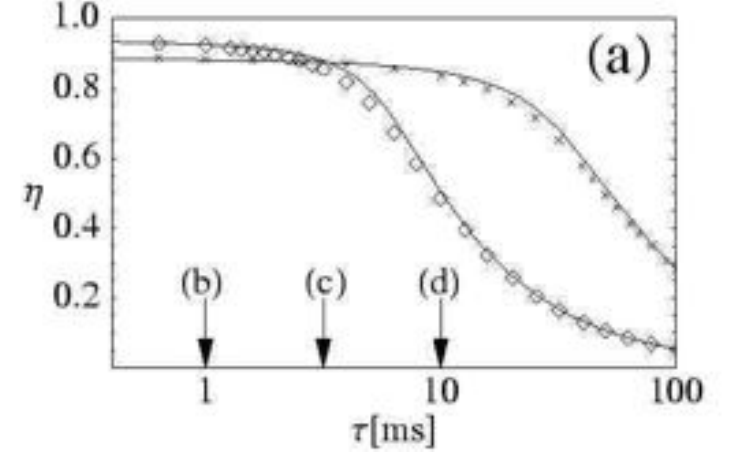}\\
\includegraphics[width=0.28\columnwidth]{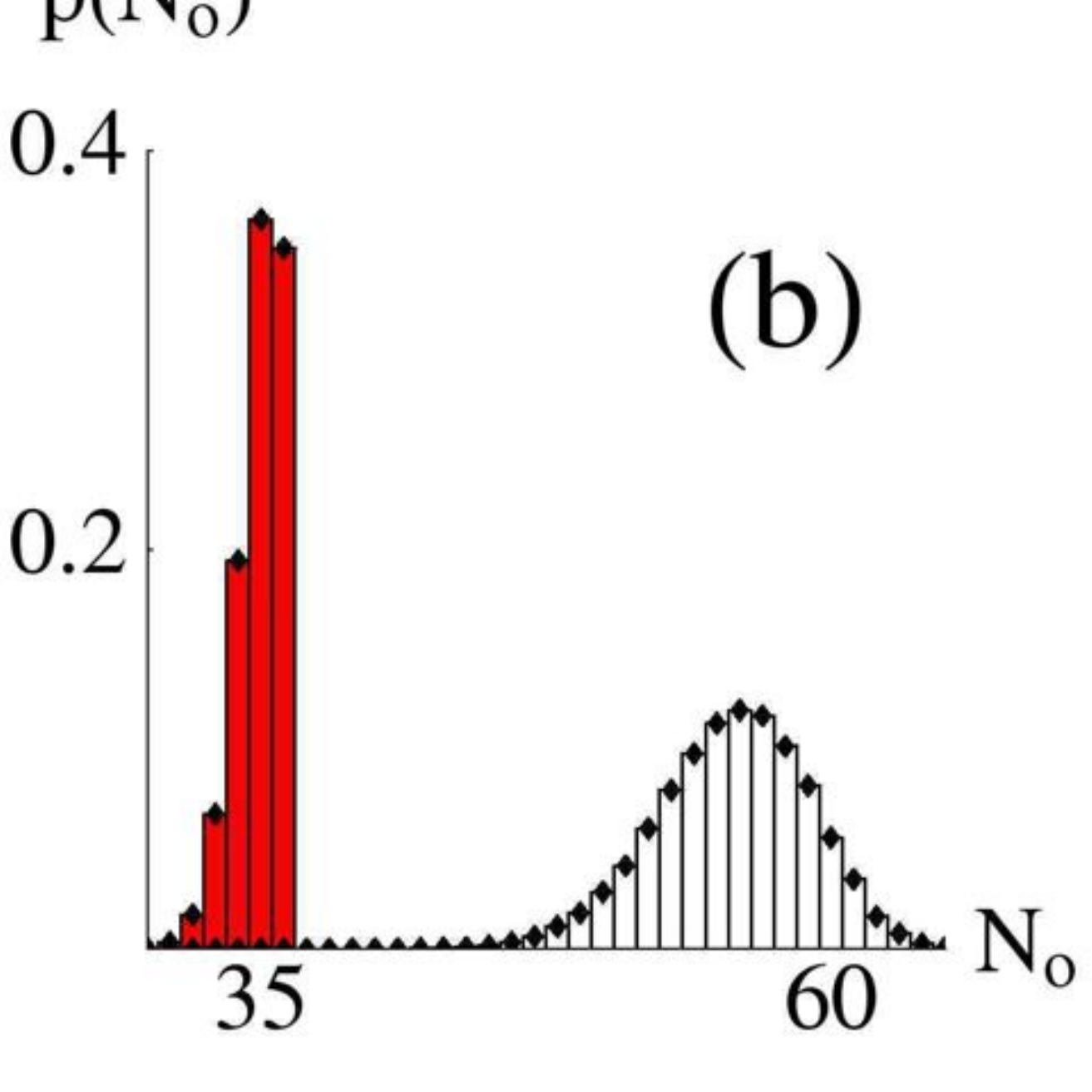}
\hspace{-0.4cm}
\includegraphics[width=0.28\columnwidth]{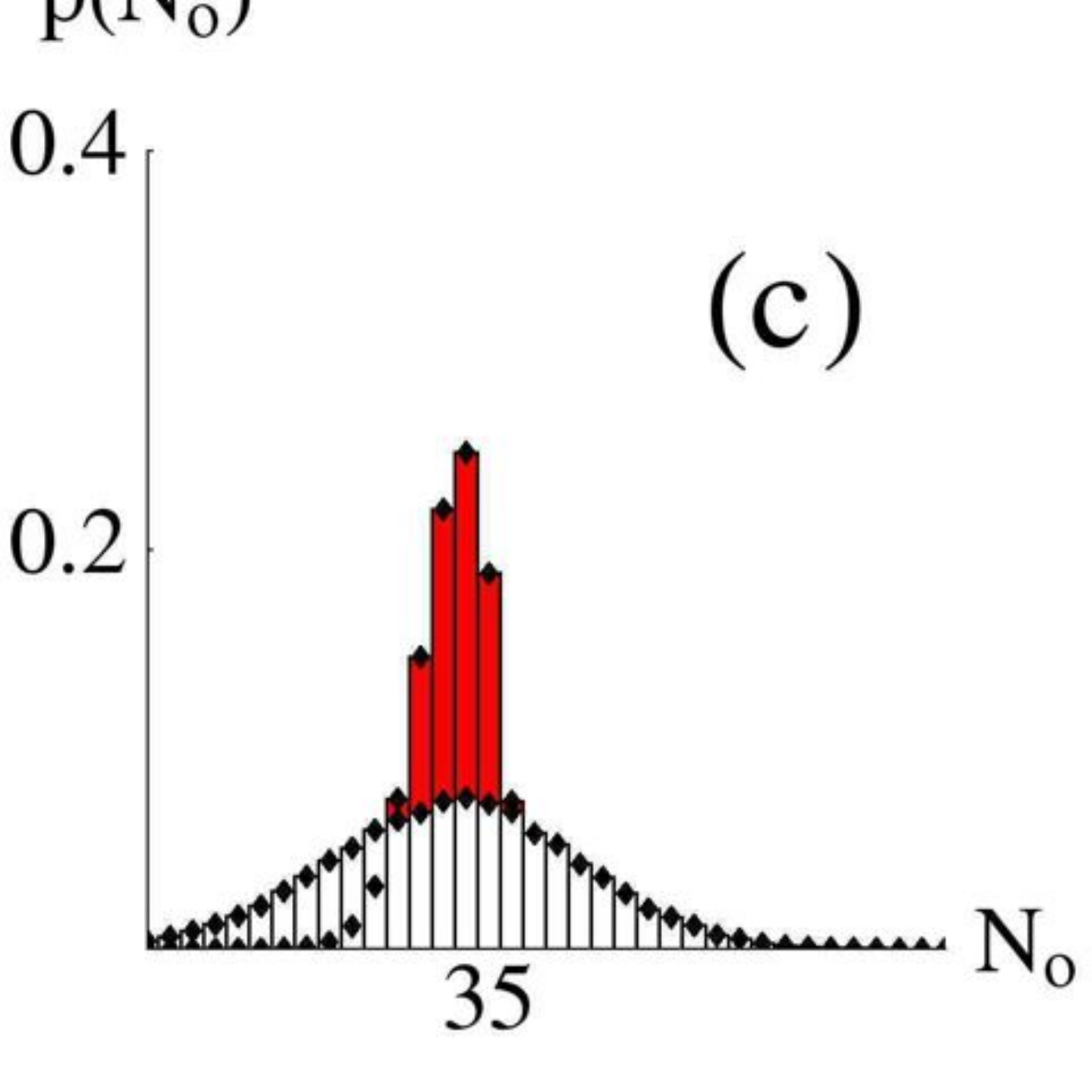}\hspace{-0.4cm}
\includegraphics[width=0.28\columnwidth]{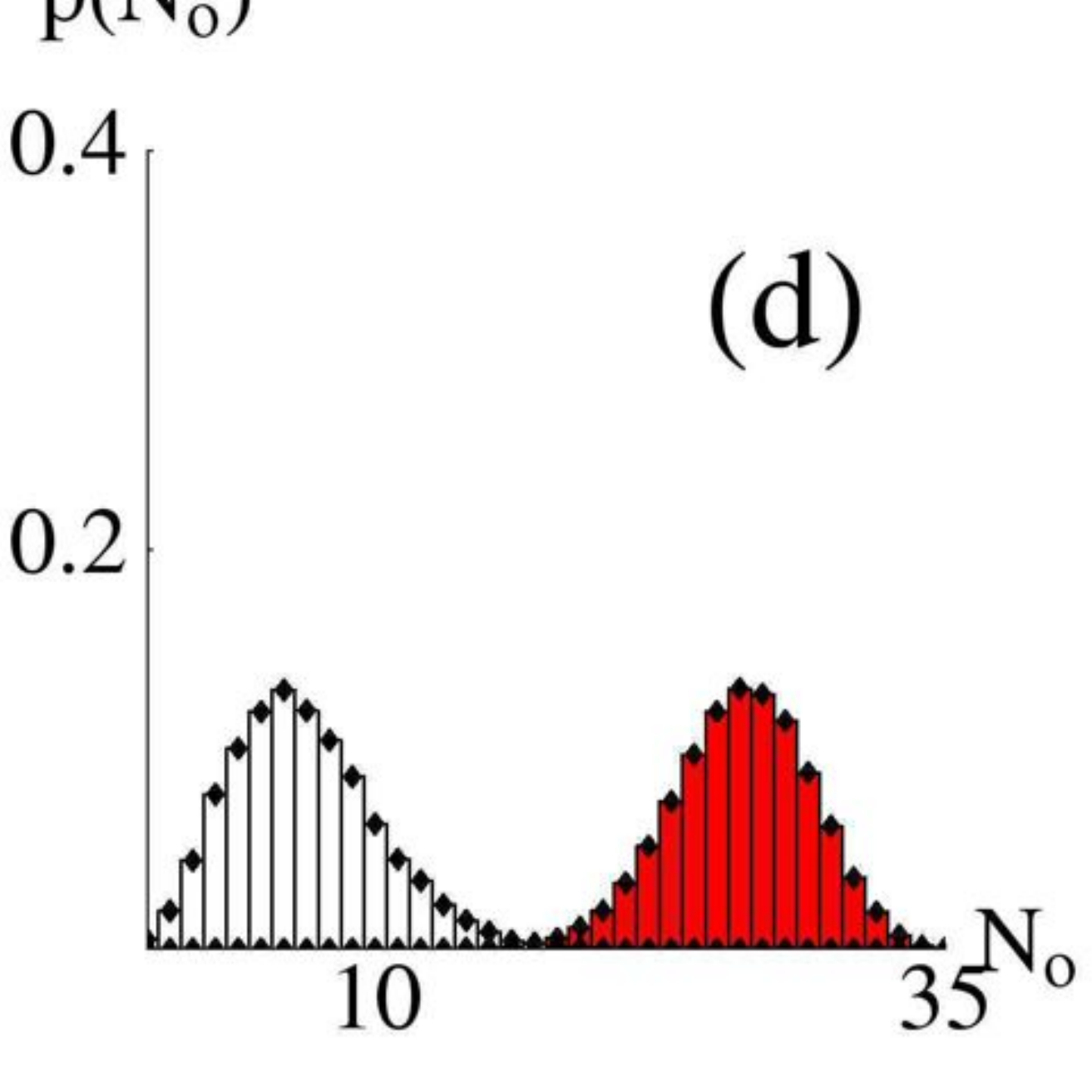}\\
\vspace{-.3cm} \caption {(a) Monte  Carlo  calculation of
efficiency $\eta$ of HLI (diamonds) and LLI (crosses) grown
membranes, as a function of the RC-cycling time $\tau$. Continuous
lines give the result of the analytical model.  (b), (c) and (d)
show  the distributions $p(N_o)$  of the number of open RCs for
the times shown with arrows in the main plot for HLI (filled bars)
and LLI (white bars).}\label{etatau}
\end{figure}

It has been here repeatedly commented that RCs perform a cycle that provides the required exciton-chemical energy conversion.  We now explain that the answer as to how adaptation can prefer the empirically observed HLI and LLI structures under different illumination conditions, lies in the
{\em interplay} between the excitation kinetics and RC cycling dynamics. By virtue of quinones-quinol and cytochrome
charge carriers, the RC dynamics features a `dead' (or
equivalently `busy') time interval during which quinol is
produced, removed and then a new quinone becomes available
\citep{savoti1997,milano2003}.  Once quinol is produced, it leaves the RC and a new quinone becomes attached. These dynamics are introduced into the simulation algorithm as presented in section \ref{simulations},   by closing an RC for a (random) poissonian distributed time  with mean $\tau$ after two excitations form quinol. The cycle can be sketched as follows: open RC$\rightarrow$ 2 ionizing excitations form quinol $\rightarrow$ closed RC in a time with mean $\tau$$\rightarrow$ open RC. This RC cycling time $\tau$ implies that at any given time, not all RCs are available for turning the electronic excitation into a useful charge separation. Therefore, the number of useful RCs decreases with increasing $\tau$. Too many excitations will rapidly close RCs, implying that any subsequently available nearby excitation will tend to wander along the membrane and eventually be dissipated - hence reducing $\eta$.
For the configurations resembling the empirical architectures
(Fig.\ref{dissmembranes}), this effect is shown  as a function of $\tau$ in Fig. \ref{etatau}(a)  yielding a wide
range of RC-cycling times at which LLI membrane is more efficient than HLI. Interestingly, this range corresponds to the measured
time-scale for $\tau$ of milliseconds \citep{milano2003}, and supports
the suggestion that bacteria improve their performance in LLI
conditions by enhancing quinone-quinol charge carrier dynamics as
opposed to manipulating exciton transfer. As mentioned, a recent proposal \citep{sturg2006} has shown numerically that the formation of LH2
para-crystalline domains produces a clustering trend of LH1
complexes with enhanced quinone availability -- a fact that would
reduce the RC cycling time.

However, the  crossover of efficiency at $\tau\approx 3$ ms  implies that even if no enhanced RC-cycling occurs, the HLI will  be less efficient than the LLI membranes on
the observed $\tau$ time-scale. The explanation is quantitatively related to the number $N_o$ of open RCs.  Figs. \ref{etatau}(b),
(c) and (d) present the distribution $p(N_o)$ of open RCs, for both HLI and LLI membranes and for the times shown with arrows in
Fig.\ref{etatau}(a).  When  the RC-cycling is of no importance (Fig. \ref{etatau}(b)) almost all RCs remain open, thereby making the HLI
membrane more efficient than LLI  since having more (open) RCs induces a higher probability for special pair oxidation. Near
the crossover in Fig. \ref{etatau}, both membranes have distributions $p(N_o)$ centered around the same value (Fig.
\ref{etatau}(c)), indicating that although more RCs are present in HLI vesicles, they are more frequently closed due to the ten fold
light intensity difference, as compared to LLI conditions. Higher values of $\tau$ (Fig. \ref{etatau}(d)) present  distributions
where the LLI has more open RCs, in order to yield a better performance when photons are scarcer. Note that distributions become wider
when RC cycling is increased, reflecting the mean-variance correspondence of Poissonian statistics used for simulation of
$\tau$.  Therefore the trade-off between RC-cycling, the actual number of RCs and the light intensity, determines the number of
open RCs and hence the performance of a given photosynthetic
vesicle architecture (i.e. HLI versus LLI).

Hence, even though these adaptations show such distinct features in the experimentally relevant regimes for the RC-cycling time and illumination intensity magnitude \citep{scheuring2004b,milano2003,savoti1997}, Figs.\ref{etatau}(c) and (d)  show that the distributions of
open RCs actually overlap implying that despite such differences in growing  environments, due to the adaptations arising, the resulting dynamics of the membranes become quite similar. Growth conditions generate  adaptations that allow on LLI membrane
 to have a larger number of open RCs than the HLI adaptation and therefore LLI membrane will perform better than HLI with respect to RC ionization irrespective of any funneling dynamics. The inclusion of RC dynamics implies that the absorbed
excitation will not find all RCs available, and somehow funneling would limit the chance of a necessary membrane sampling to explore further open RCs. Globally, a given amount of closed RCs  will eventually alter the excitation's fate since
probable states of oxidization are readily reduced. In a given
lifetime, an excitation will find (depending on $\tau$ and current light intensity $I$) a
number of available RCs -- the {\it effective
stoichiometry} -- which is different from the actual number reported by
Atomic Force Microscopy \citep{scheuring2004b,bahatyrova2004}.

The effect of incident light intensity variations relative to the light intensity during
growth with both membranes, presents a similar behavior. In Ref. \citep{NJPCaycedo-Soler,PRLCaycedo-Soler} such study is performed and it is concluded that: the LLI membrane performance starts to diminish well beyond the growth light intensity, while the HLI
adaptation starts diminishing just above its growth intensity due to rapid RC closing that induce increased
dissipation. Hence,  in LLI membranes excess
photons are readily used for bacterial metabolism, and HLI
membranes exploit dissipation in order to limit the number of
processed excitations. In the same work, it is found that the effect of the arrangement itself is lost due
to RC dynamics, since the effective stoichiometry with spread out open RCs becomes alike among different membranes  sharing the same amount of RCs, equal cycling time $\tau$ and incident light intensity.

To summarize so far,  the arrangement of complexes changes slightly the efficiency of the membranes when no RC dynamics is included -- but with RC dynamics, the most important feature is the number of open RCs. Although the longer RC closing times make membranes more prone to dissipation and decreased efficiency,
it also makes the architecture less relevant for the overall dynamics.  The relevant network architecture instead becomes the dynamical one including sparse  open RCs, not the static geometrical one involving the actual clustered RCs.

\section*{4. Analytical model}\label{analyt}
Within a typical fluorescence lifetime of 1 ns, a single excitation has travelled hundreds of sites and explored the available RCs globally. The actual arrangement or architecture of the complexes seems not to influence the excitation's fate, since the light intensity and  RC cycling determine the number of open RCs and the availability for P oxidation.

\subsection*{4.1 Excitation transfer-RC cycling rate  model}
Here, we present an alternative rate model which is inspired by the findings of the numerical simulations, but which (1)
globally describes the excitation dynamics and RC cycling, (2)
leads to analytical expressions for the efficiency of the membrane
and the rate of quinol production, and (3) sheds light on the
trade-off between RC-cycling and exciton dynamics \citep{PRLCaycedo-Soler,NJPCaycedo-Soler}.

\begin{figure}
\centering
\includegraphics[width=0.45\columnwidth]{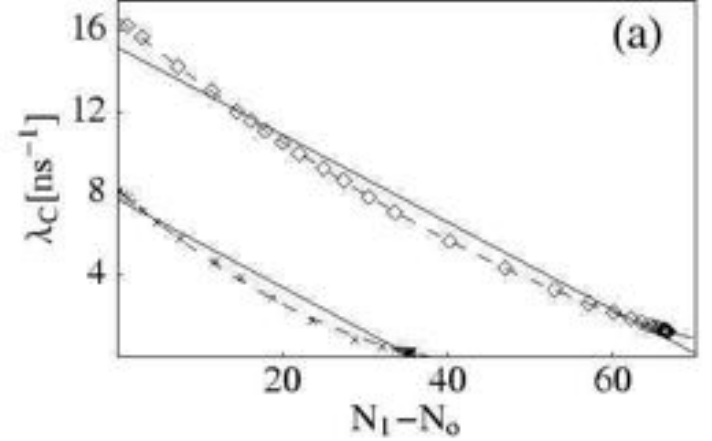}
\includegraphics[width=0.45\columnwidth]{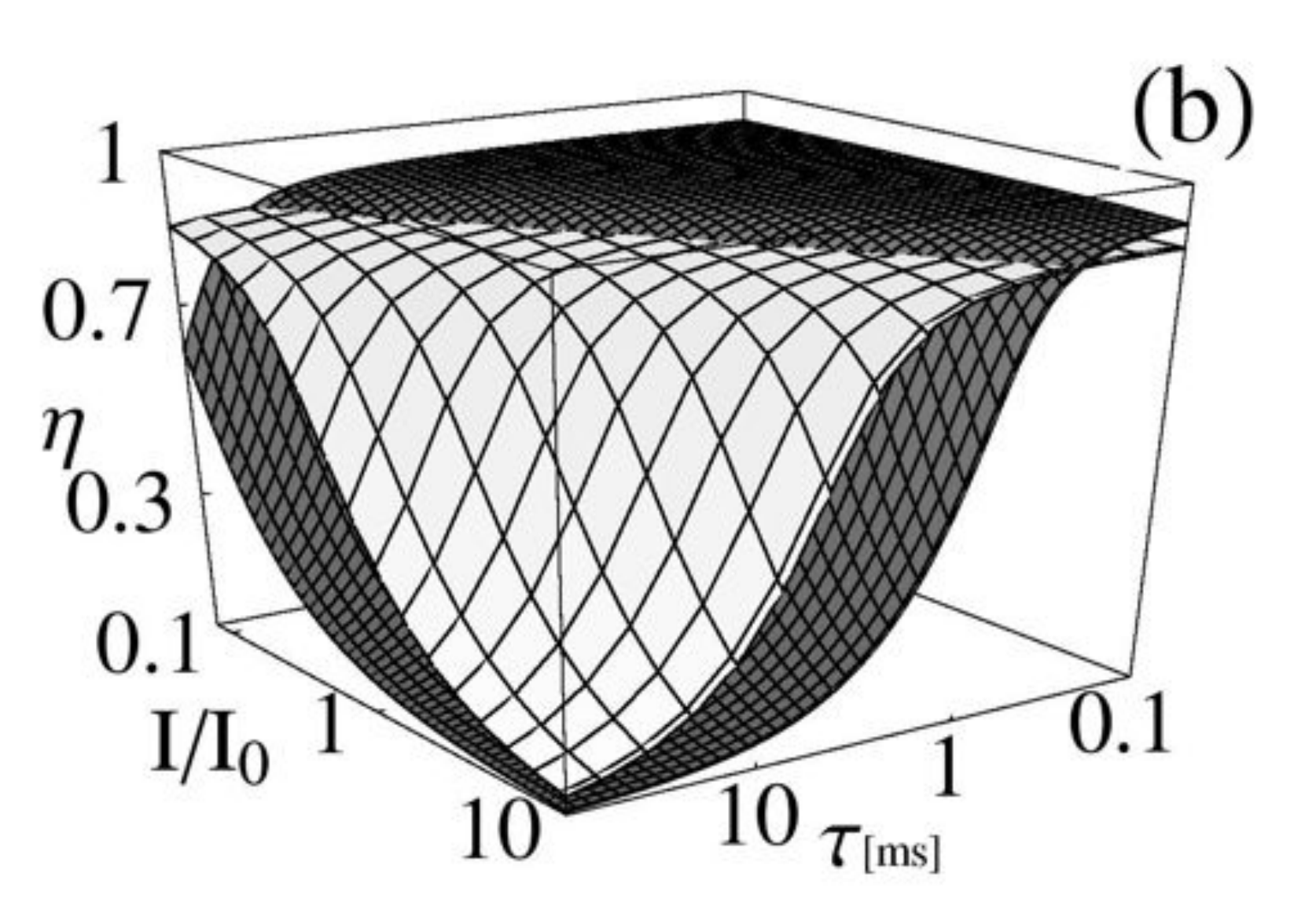}\\
\caption{(a) Numerical results showing the rate of ionization $\lambda_C(N_o)$ of an RC for HLI (diamonds) and LLI (crosses) membranes, together with a quadratic (dashed line) and linear (continuous) dependence on the number of closed RCs $(N_{1}-N_{o})$. The fitting parameters for $a+bN_o$ are $a=\{15.16, 7.72\} $ns$^{-1}$, $b=\{-0.21,-0.21\}$ ns$^{-1}$; and for  $a+bN_o+cN_o^2$, $a=\{16.61,8.21\}$ns$^{-1}$, $b=\{-0.35,-0.33\}$, and $c=\{3.6, 1.5\} \mu$s$^{-1}$, for HLI and LLI membranes respectively. (b) $\eta$ as function of $\tau$ and $\alpha=I/I_0$, obtained from the complete analytical solution for LLI (white) and HLI (grey) membranes}\label{eta3D}
\end{figure}

Shortly, $N_E$ excitations area absorbed by the membrane at a rate $\gamma_A$, and will find its way to become RC ionizations with a rate per particle $\lambda_C(N_o)$ whose dependence on the number of open RCs is made explicit, or be dissipated at a rate $\gamma_D$. On the other hand, RCs have their own dynamycs, closing at a rate $\lambda_C(N_o)\times N_E$ and individually opening at a rate $1/\tau$. The dependance   $\lambda_C(N_o)$  is numerically found aidded by the stochastic model and shown in Fig.\ref{eta3D}(a). Rate equations can therefore be written:
\begin{eqnarray}
\frac{dN_E}{dt}&=&-(\lambda_C(N_o)+\gamma_D) N_E+\gamma_{A}  \label{ne}\\
\frac{dN_{o}}{dt}&=&\frac{1}{\tau}(N_{1}-N_{o})-\frac{\lambda_C(N_o)}{2} N_E. \label{noff}
\end{eqnarray}
to be solved, and of use for the calculation of the steady-state efficiency $\eta=n_{RC}/n_A$:
\begin{eqnarray}
\eta=\frac{\lambda_C(N_{o}) N_E}{\gamma_A}.
\end{eqnarray}
A linear fit for
$\lambda_C(N_o)$  allows an analytical expression for $\eta$:
\begin{eqnarray}
 \eta(\tau,\gamma_A(I))=\frac{1}{\gamma_A\lambda_C^0\tau}\left\{2N_1(\lambda_C^0+\gamma_D)+\gamma_A\lambda_C^0\tau-\right.& &\\
\left.\sqrt{4N_{1}^2(\lambda_C^0+\gamma_D)^2+4 N_{1}\gamma_A\lambda_C^0(\gamma_D-\lambda_C^0)\tau+(\gamma_A\lambda_C^0\tau)^2}\right\}\ \
\end{eqnarray}
where $\lambda_C^0$ is the rate of RC ionization when no RC-cycling is accounted, dependent only on the amount of  RCs present in the vesicle \citep{NJPCaycedo-Soler}. This analytical expression is shown in  Fig.\ref{eta3D}(b) and  illustrates that
$\eta \geq 0.9$ if the transfer-P reduction time  is less than a
tenth of the dissipation time, not including RC cycling. As
can be seen in  Figs. \ref{etatau}(a), the analytical
solution is in good   quantitative agreement with the numerical
stochastic simulation, and provides support for the assumptions made. Moreover, this model shows directly that the efficiency is driven by
the interplay between the RC cycling time and light intensity.
Figure \ref{eta3D}(b) shows up an entire region of parameter space
where LLI membranes are better than HLI in terms of causing P
ionization, even though the actual number of RCs that they have is
smaller. In view of these results, it is interesting to note how clever Nature has been in tinkering with the efficiency of LLI vesicles and the dissipative behavior of
HLI adaptation, in order to meet the needs of bacteria
subject to the illumination conditions of the growing environment.

\subsection*{4.2 Bacterial metabolic demands}

\begin{figure}
\centering
\includegraphics[width=0.47\columnwidth]{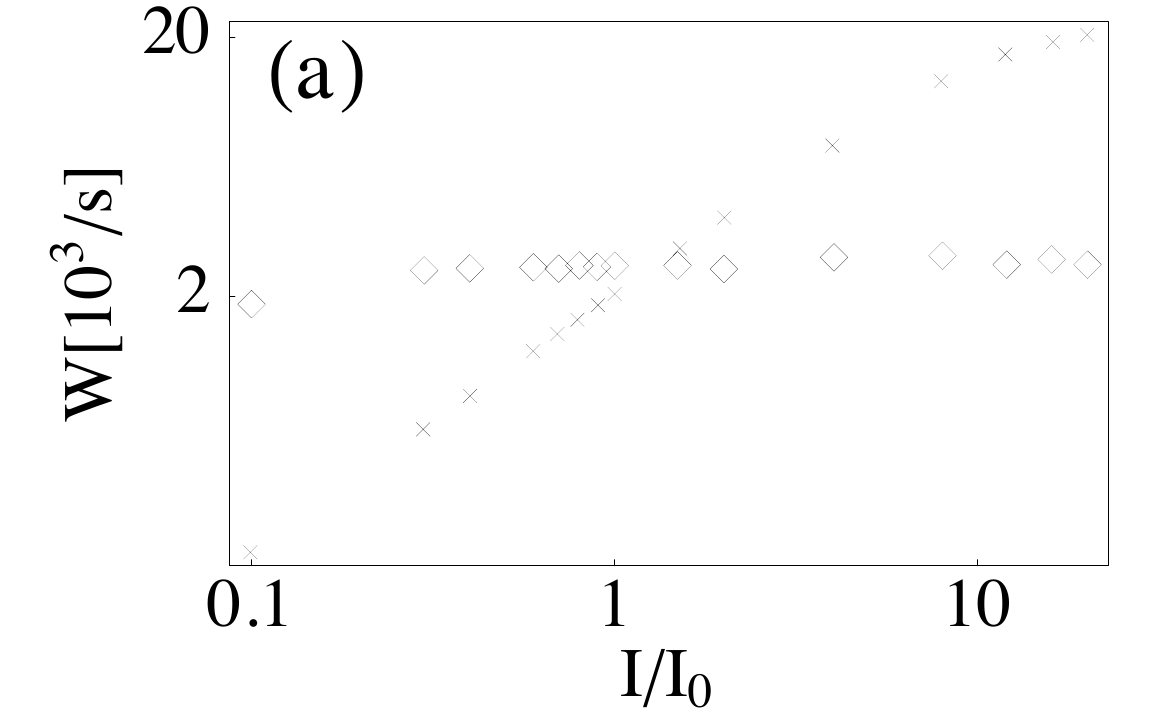}
\includegraphics[width=0.5\columnwidth]{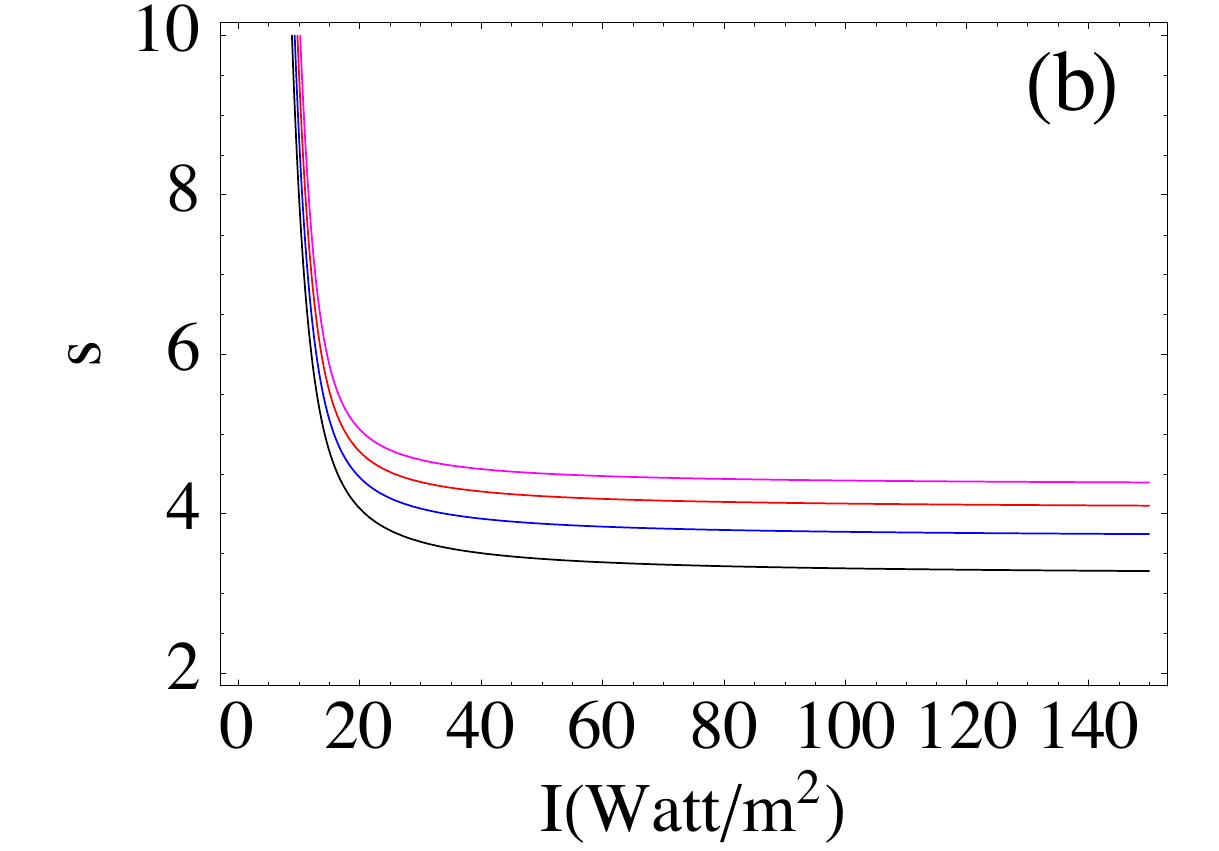}
\caption{  (a) Quinol rate $W$ in HLI (diamonds, $I_0=100$W/m$^2$)
and LLI (crosses, $I_0=10$W/m$^2$) grown membranes, as a function
of incident intensity $I$. (b) Quinol rate contours of $W=\{1900,2000,2100,2200\}$ s$^{-1}$ in black, blue, red and pink, respectively.}\label{spred}
\end{figure}

Photosynthetic membranes must provide enough energy to fulfill the
metabolic requirements of the living bacteria quantified by the
quinol output or quinol rate
\begin{equation}
W=\frac{1}{2}\frac{dn_{RC}}{dt}
\end{equation}
which depends directly on the excitations that ionize RCs $n_{RC}$. The factor $\frac 12$ accounts for the requirement of two
ionizations to form a single quinol molecule.
Although these membranes were
grown under continuous illumination, the adaptations themselves
are a product of millions of years of evolution. Using RC cycling
times that preserve quinol rate in both adaptations, different
behaviors emerge when the illumination intensity is varied (see
Fig. \ref{spred}(a). The increased illumination is readily  used by
the LLI adaptation, in order to profit from excess excitations in
an otherwise low productivity regime. On the other hand, the HLI
membrane maintains the quinone rate constant, thereby avoiding the
risk of pH imbalance in the event that the light intensity
suddenly increased. We stress that the number of RCs synthesized does not
directly reflect the number of available states of ionization in
the membrane. LLI synthesizes a small amount of RCs in order
to enhance quinone diffusion, such that excess light intensity is
utilized by the majority of special pairs. In HLI, the synthesis
of  more LH1-RC complexes  slows down RC-cycling, which ensures
that many of these RCs are unavailable  and hence be advantageous of evenly distributed dissipation to steadily supply quinol
  independent of any excitation increase. The
very good agreement between our analytic results and the stochastic
simulations, yields additional physical insight concerning the
stoichiometries found experimentally in {\it Rsp.
Photometricum}.

A closed form expression regarding all dynamical parameters involved can be obtained \citep{NJPCaycedo-Soler}:
\noindent{\small{\small\begin{equation}
2W(s,I)=\frac{\gamma_A(s,I)}{2}+ \frac{1}{B(s)}
\left(1+\frac{\gamma_{D}}{\lambda_c^0}\right)
+
\sqrt{[\frac{\gamma_A(s,I)}{2}+\frac{1}{B(s)}\left(1+\frac{\gamma_{D}}{\lambda_c^0}\right)]^2+\frac{\gamma_A(s,I)}{2B(s)}}
\end{equation}}}
\noindent where the dependence on stoichiometry is made explicit due to absoprtion cross section in $\gamma_A$ and on $B(s)=\frac{\tau(s) (A_1 +s A_2)}{f(s)A_0}$, which is  a parameter that depends on area $A_{1(2)}$ of individual LH1 (LH2) complexes and filling fraction $f(s)$. The filling chromatophore fraction dependence on $s$  is available from experimental data of Ref.\citep{sturg2006} and $\tau(s)$ is constructed from an interpolation scheme \citep{NJPCaycedo-Soler}.

As emphasized in Ref.  \citep{scheuring2005},  membranes
with $s$=6 or $s$=2 were not observed, which is to be compared with the contour plots regarding constant quinol output, of $s$ as a function of growing light intensity $I_0$, shown in Fig.\ref{spred}(b). These results support a dichotomic observation where $s\approx 4$  predominantly on a  great range for growing light intensity. However, a prediction can be made for 30-40 W/m$^2$ where a great sensitivity of stoichiometry ratios rapidly build up the number of antenna LH2 complexes. Very
recently \citep{sturg2009}, membranes were grown with 30W/m$^2$ and an experimental stoichiometry of 4.8 was found. The contour of 2200 s$^{-1}$ predicts a value for stoichiometry of 4.72 at such growing light intensities. This agreement is quite remarkable, since a
simple linear interpolation among the values $s\approx 4, I_0=100$ W/m$^2$ and $s\approx 8, I_0=10$ W/m$^2$ would  wrongly predict $s=7.1$ at 30 W/m$^2$. We encourage experimentalists to confirm the full range of predicted behaviors as a function of light-intensity and stoichiometry. Such exercise would without doubt confirm/expand the understanding on RC-exciton dynamics trade off,  pinpointing a direction to pursue  solar energy conversion research, provided by Nature itself.

\section*{4.3 Performance of photosynthetic membranes under extreme photon statistics}\label{BM}

Photosynthetic (e.g purple) bacteria provide the crucial coupling between the Sun's energy and the production of food on Earth, and have adapted successfully to a variety of terrestrial conditions since the beginnings of life on Earth several billion years ago. In this section we explore whether terrestrial bacteria, which are the product of millions of years of evolutionary pressure on Earth, could survive if suddenly exposed to incident light with extreme statistics. We are therefore mimicking a scenario in which purple bacteria are either (i) suddenly  transported to some unknown extreme solar environment elsewhere in the universe, or (ii) where our own Sun suddenly picks up extremal behavior in terms of the temporal statistics of its emitted photons, or (iii) the bacteria are subjected to extreme artificial light sources such as that in Ref.
\citep{Jalali2009} involving stimulated Raman scattering, as well as in
coherent anti-Stokes Raman scattering in silicon, or (iv) the external membrane skin of the bacterium is modified in such a way that the absorption of photons takes on extreme statistical properties.

Our discussion is qualitatively different from previous work looking at life in extreme conditions, since those discussions have tended to focus on {\em environmental} extremes affecting the  biochemical metabolism or extremes in the incident light intensity. By contrast, our current discussion focuses on extremes in the incident photon statistics.
Although the setting is largely hypothetical, our theoretical predictions are based on a realistic semi-empirical model which incorporates (i) high precision empirical AFM information about spatial locations of LH1 and LH2 biomolecular complexes in the membrane architecture of {\it Rsp. Photometricum} purple bacteria, and (ii) full-scale stochastic simulations of the excitation kinetics and reaction center dynamics within the empirical membrane.

As with any process involving events occurring in a stochastic way over time, the statistical properties of arriving photons may show deviations from a pure coin-toss process in two broad ways: burstiness and memory \citep{Barabasi2008}. First consider the simplest process in which the rate of arrival of a photon has a constant probability per unit time. It is well known that this so-called Poisson process produces a distribution of the waiting time for the next photon arrival which is exponential in form, given by $P_{\rm P}(\tau)\sim {\rm exp}(-\tau/\tau_0)$. The extent to which the observed arrival time distribution $P(\tau)$ deviates from exponential, indicates how non-Poissonian the photon arrival is. Following Barabasi \citep{Barabasi2008}, we refer to this as `burstiness' $B$ and define it by its deviation from a purely Poisson process:

$$ B\equiv \frac{(\sigma_\tau/m_\tau-1)}{(\sigma_\tau/m_\tau+1)}=
\frac{(\sigma_\tau -m_\tau)}{(\sigma_\tau+m_\tau)} $$

\noindent where $\sigma_\tau$ and $m_\tau$ are the standard deviation and mean respectively of the empirical distribution $P(\tau)$. For a pure Poisson process, the mean and standard deviation of the arrival time distribution are equal and hence $B=0$. The other property which can be noticeable for a non-Poisson process is the memory $M$ between consecutive inter-arrival times which, following Barabasi, we define as:

$$ M\equiv \frac{1}{n_\tau -1} \sum_{i=1}^{n_\tau-1}
\frac{(\tau_i-m_1)(\tau_{i+1}-m_2)}{\sigma_1 \sigma_2} $$

\noindent where $n_\tau$ is the number of inter-arrival times measured from the signal and $m_1(m_2)$ and $\sigma_1(\sigma_2)$ are sample mean and sample standard deviation of $\tau_i (\tau_{i+1})$'s respectively ($i=1,\dots,n_\tau-1$). For a pure Poisson process, $M=0$.
Both $B$ and $M$ range from $-1$ to $+1$, with the value $B=0$ and $M=0$ for a strict Poisson process. We will assume for simplicity that the arriving photons are all absorbed and hence one exciton created within the membrane by each photon. This can easily be generalized but at the expense of adding another layer of statistical analysis to connect the statistics of arriving photons to the statistics of the excitons being created within  the LH2/LH1 membrane -- indeed, taking a constant absorption probability less than unity would not change our main conclusions. We will also neglect the possibility that several photons arrive at exactly the same time. In principle, incident photons can be generated numerically with values of $B$ and $M$ which are arbitrarily close to any specific $(M,B)$ value -- however this is extremely time-consuming numerically. Instead, we focus here on specific processes where the $B$ and $M$ can be calculated analytically. Although this means that the entire $(M,B)$ parameter space is not accessed, most of it can indeed be -- and with the added advantage that analytic values for $B$ and $M$ are generated.

\begin{figure}[htb]
\centering
\includegraphics[width=1\columnwidth]{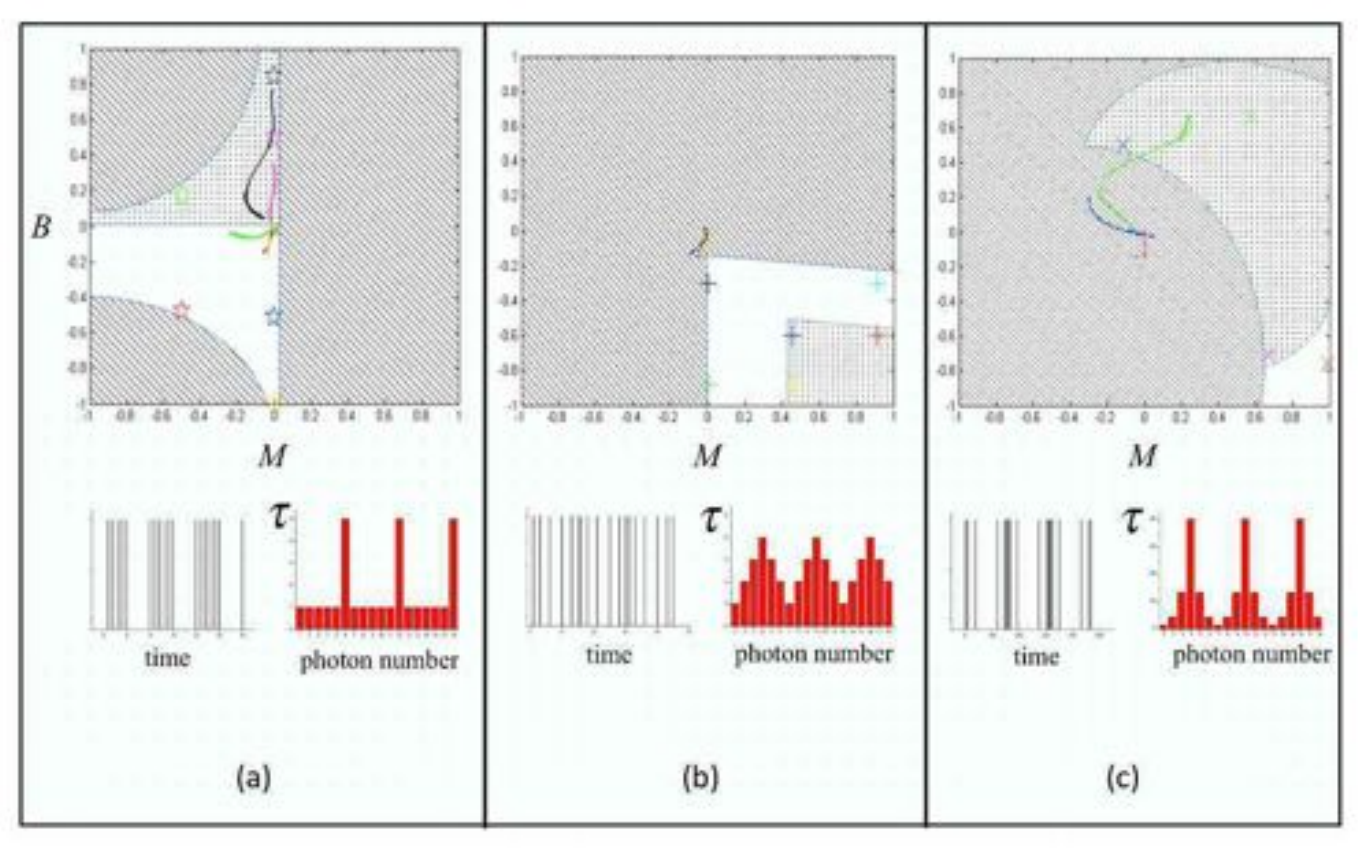}
\caption{The $(M,B)$ phase diagram governing survivability of a terrestrial bacteria under a wide range of incident photon time-series with a given burstiness $B$ and memory $M$. The sum of the blank and dotted areas in each diagram is the region accessible by each of the three photon processes shown. For each $(M,B)$ value, a time-series of photons is created with these properties and used to generate excitations within the membrane. The resulting quinol output time-series is then calculated. Assuming that this output needs to be similar to that on Earth, the dotted region corresponds to photon $(M,B)$ input values for which the bacteria would die, while the blank region corresponds to photon input for which the bacteria would survive. The trajectories represent the range of quinol outputs from the reaction centers (RC) as we span the range of physically reasonable RC closure times, given a particular input (shown as a star or cross of the same color). As the time during which the RC is closed increases, the $B$ and $M$ output values tend to move toward the origin (i.e. toward $B=0$ and $M=0$). Bottom: Photon inputs correspond to (a) step input, (b) bunched input and (c) power-law step input. Lower row in each case shows photon arrival process (left, black barcode) and waiting time $\tau$ between photon arrivals (right, red histogram).} \label{BMfig}
\end{figure}

Figure \ref{BMfig} summarizes our findings in terms of the incident photons (i.e. excitation input) and metabolic output from the LH2/LH1 membrane, over the entire $(M,B)$ parameter space. The subspace shown by the combination of the white and dotted regions (i.e. the region in the $(M,B)$ space which is {\em not} diagonally shaded) comprises points with $(M,B)$ values that can be generated by one of three relatively simple types of photon input: (a) step input, (b) bunched input and (c) power-law step input, as shown in the three panels respectively. Each point in the blank or dotted region denotes a time-series of initial excitations in the membrane with those particular burstiness and memory values ($B$ and $M$).  This train of initial excitations then migrates within the membrane of LH2/LH1 complexes, subject to the dynamical interplay of migration and trapping as discussed earlier in this paper, and gives rise to a given output time-series of 'food' (quinol) to the bacteria. Figure \ref{BMfig} shows explicitly a variety of initial $(M,B)$ values (crosses and stars) and the trajectory represents the locus of resulting quinol outputs from the reaction centers (RC) for this particular $(M,B)$ input. The trajectory is generated by varying the RC closing time within the range of physically reasonable values. The trajectory is finite since the range of physically reasonable RC closing times is also finite (20 to 1000Hz). As the time during which the RC is closed increases, the output becomes more Poisson-like, i.e. the $B$ and $M$ output values from the membrane after absorption at the RC, tend towards 0. Hence the trajectories head toward the center as shown.

The $(M,B)$ values for natural sunlight on Earth lie near $M=0$ and $B=0$ since the incident photon arrival from the Sun is approximately a Poisson process. This produces a quinol output which is also a Poisson process ($M=0$ and $B=0$). Now suppose a terrestrial bacteria is suddenly subjected to an extreme incident light source with $(M,B)$ values which lie at a general point in $(M,B)$ parameter space. Can it survive? It is reasonable to expect that the
bacteria will not survive if the resulting quinol output is very different from that on Earth (i.e. very different from $B=0$ and $M=0$) since the bacteria is well-tuned for Life on Earth only. Hence we say that the bacteria can only survive if the quinol output has $(M,B)$ values within 0.01 of the terrestrial values of $B=0$ and $M=0$. The blank spaces denote $(M,B)$ values for which the bacteria would survive, while the dotted spaces are where the bacteria would die.

Remarkably, there are therefore substantial regions of extreme incident photon statistics (i.e. $B\neq 0$ and $M\neq 0$) where the bacteria would survive, producing a quinol output for the rest of the organism which resembles that on Earth (i.e. $|B|<\epsilon$ and $|M|<\epsilon$ where $\epsilon=0.01$). The fact that these blank spaces have  an irregular form is due to the highly nonlinear nature of the kinetic interplay between exciton migration and trapping, as discussed earlier in this chapter.
For example, panel (c) shows that a bacterium which is only tuned to survive on Earth can have a maximally negative burstiness $B=-1$ and a maximal memory $M=1$, meaning that the incident photons arrive like the regular ticks of a clock. The same applies to all the incident photon conditions demonstrated by the blank spaces. If we make the survival criterion more generous (i.e. $\epsilon>0.01$) the blank spaces corresponding to survivability of the terrestrial bacterium increase in size until no dotted space remains (i.e. when $\epsilon=1$). All the dotted areas in Fig. \ref{BMfig} become blank spaces and hence represent survivable incident photon conditions for the bacteria.

\section*{5. Perspectives: Photosynthetic membranes of purple bacteria as a basis to develop stable energy conversion}

This chapter has covered a comprehensive review of purple bacteria adaptation to light intensity conditions and has provided a basis to understand the chromatic adaptation of photosynthetic harvesting membranes as a consequence of the requirement imposed by nature on living organisms to develop a machinery capable of fuel production in a stable manner. It is worth to emulate, and certainly interesting to develop in artificial devices to surpass the usual bleaching on semiconductor panels, an analogous mechanism to the excitation kinetics and RC dynamics interplay, in order to shift the emphasis of requirements of fuel production and allow highly efficient energy transfer to available charge carriers (quinones) in low light intensity conditions, while consistently, a leveled fuel (quinol) production statistics is accomplished when photon arrival varies both in intensity and photon waiting time statistics.

\section*{Acknowledgments}
The authors acknowledges financial support from Research Projects Facultad de Ciencias, Universidad de los Andes, Banco
de la Republica and Fundacion Mazda.

\end{document}